\documentclass[aps,pra,reprint,onecolumn,11pt,tightenlines,raggedbottom,notitlepage,longbibliography,nofootinbib]{revtex4-2}
\usepackage[T1]{fontenc}
\usepackage{lmodern}
\usepackage{helvet}
\usepackage{amsmath,amssymb,amsthm,bm,mathtools}
\usepackage{microtype}
\usepackage{placeins}
\usepackage{graphicx}
\usepackage[hidelinks]{hyperref}
\hypersetup{pdftitle={Boson sampling is fragile to loss},pdfauthor={Youngrong Lim}}
\newtheorem{theorem}{Theorem}
\newtheorem{lemma}{Lemma}
\newtheorem{corollary}{Corollary}
\theoremstyle{definition}
\newtheorem{problem}{Problem}
\newcommand{\E}{\mathbb E}
\newcommand{\TV}{d_{\mathrm{TV}}}
\newcommand{\Tr}{\operatorname{Tr}}
\newcommand{\Sym}{\operatorname{Sym}}

\newcommand{\HS}{\mathrm{HS}}

\newcommand{\clip}{\operatorname{clip}}
\newcommand{\sech}{\operatorname{sech}}
\newcommand{\ket}[1]{|#1\rangle}
\newcommand{\bra}[1]{\langle#1|}

\begin{document}
\title{Boson sampling is fragile to loss}
\author{Youngrong Lim}
\email{sshaep@gmail.com}
\affiliation{Department of Physics, Chungbuk National University, Cheongju 28644, Korea}

\begin{abstract}
Boson sampling experiments aim to demonstrate quantum computational advantage with increasing numbers of interfering photons. A central question is whether their computational hardness survives under photon loss. We show that even an arbitrarily small nonzero uniform loss rate permits efficient classical sampling in the average case when the rate remains fixed as the system grows. We exploit how loss suppresses correlations involving many photons to approximate the full output distribution using low correlation orders. The resulting samplers run in time polynomial in system size and inverse total variation error, with the error averaged over Haar random interferometers. The results apply to Fock state boson sampling with single-photon states in distinct input modes and arbitrary additional vacuum inputs, and to Gaussian boson sampling with equal squeezing of bounded strength in every input mode. Classical sampling is efficient even when the mean detected photon number grows linearly with system size, showing that large photon counts alone do not ensure computational hardness.
\end{abstract}
\maketitle

\section{Introduction}\label{sec:introduction}

In an optical boson sampler, indistinguishable photons interfere in a network of beam splitters and phase shifters, and detectors record the photon number at each output. The computational task is to sample this joint output distribution for a specified input state and interferometer. Experimental runs provide these samples directly, whereas classical simulation is believed to be intractable under plausible assumptions~\cite{AaronsonArkhipov}. This makes boson sampling a candidate for demonstrating quantum computational advantage without a universal quantum computer. Photonic implementations toward this goal use single photons~\cite{Ascella2024} or squeezed light~\cite{Zhong2020,ArrazolaChip2021,Jiuzhang2021,Borealis2022,Jiuzhang2023,Jiuzhang2026,Clavina2026} in free-space optics, optical fibers and integrated circuits, with increasing numbers of photons and optical modes.

Photon loss is a severe parameter in experiments. Jiuzhang 4.0, for example, reports an overall system efficiency of 51\%, corresponding to detection of only about half the generated photons~\cite{Jiuzhang2026}. Assessing quantum computational advantage in these experiments therefore requires understanding the classical cost of sampling from their lossy output distributions. Conditional hardness has been established when only a logarithmic number of photons is lost on average~\cite{AaronsonBrod,GoDistinguishability,GoLoss}, whereas efficient classical simulation is known under sufficiently strong loss that the surviving fraction tends to zero~\cite{OszmaniecBrod2018,GarciaPatron2019,Qi,OhNP,LimReset,OhMPS2026}. These earlier results left a broad range between the two regimes unresolved, including the experimentally relevant case of a fixed loss rate. At a fixed rate, both the mean detected count and the mean number of lost photons grow in proportion to the input size.

For two standard families of input states, we show that a classical computer can efficiently sample the lossy output distribution at any fixed rate of uniform photon loss, however small. The first family is Fock state boson sampling with one photon in each occupied input mode and any number of empty modes. The second is Gaussian boson sampling with all input modes squeezed equally, by an amount that stays bounded as the system grows. Accuracy is the total variation distance between the distribution produced by the classical sampler and the lossy output distribution, taken over all output configurations and averaged over Haar random interferometers. Bringing this distance below any target takes time polynomial in the number of modes and in the inverse of the target. The running time therefore stays polynomial even when the error shrinks as an inverse polynomial in system size.

These are exactly the conditions under which the hardness of boson sampling is argued, with the same input states, Haar random interferometers and inverse polynomial total variation error~\cite{AaronsonArkhipov,Hamilton,GoLoss}. Our results differ from those arguments only in the amount of photon loss. In the known hard regime, only a logarithmic number of photons is lost on average, whereas we prove efficient sampling whenever a fixed fraction of the photons is lost. The unresolved range between the two regimes is thereby narrowed to loss rates that decrease toward zero as the system grows. Experiments performed with interferometers close to Haar random at a fixed loss rate therefore cannot guarantee quantum computational advantage.

\section{Problem}\label{sec:problem}

In a boson sampling task, an input state of $M$ optical modes passes through a passive linear interferometer, and photon number resolving detectors count the photons in every output mode. The interferometer acts on the mode operators through an $M\times M$ unitary matrix $U$. Each photon reaches the detectors with probability $\eta$, independently of the other photons and of its path through the interferometer (Fig.~\ref{fig:mechanism}a). We call $0<\eta<1$ the transmission and $1-\eta$ the loss rate. A run yields an output configuration $x=(x_1,\ldots,x_M)$ with $k=\sum_ix_i$ detected photons, and $P_U(x)$ denotes its probability for a given input state. This distribution includes every value of $k$ and every configuration with several photons in one output mode. Although we use optical language, the analysis relies only on bosonic statistics and uniform particle loss, so it applies equally to other noninteracting bosons, such as atoms in optical lattices~\cite{Atomic2024}.

The interferometer is drawn once from the Haar measure, the uniform distribution on $M\times M$ unitary matrices, and is then kept fixed for all runs. A classical sampler receives $U$ and outputs samples from a distribution $Q_U$. We measure its error by the total variation distance $\TV(P_U,Q_U)=\frac12\sum_x|P_U(x)-Q_U(x)|$, which bounds the error in the probability of every set of output configurations. We consider two input families.

\begin{problem}[Lossy Fock state boson sampling]\label{prob:fock}
Single photons occupy $n\le M$ distinct input modes, and the remaining modes are in vacuum,
\begin{equation}
 \ket{\Psi_{\mathrm F}}=\ket1^{\otimes n}\ket0^{\otimes(M-n)}.
 \label{eq:fock-input}
\end{equation}
Given $U$, $\eta$ and $\epsilon>0$, generate samples from a distribution $Q_U$ with $\E_{U\sim\mathrm{Haar}}\TV(P_U,Q_U)\le\epsilon$, where $P_U$ is the output distribution of this input under uniform loss with transmission $\eta$.
\end{problem}

\begin{problem}[Lossy Gaussian boson sampling]\label{prob:gaussian}
Squeezed vacuum states with a common squeezing parameter $r\ge0$ occupy $s\le M$ input modes, and the remaining modes are in vacuum,
\begin{equation}
 \ket{\Psi_{\mathrm G}}=\ket{\zeta_r}^{\otimes s}\ket0^{\otimes(M-s)},\qquad
 \ket{\zeta_r}=\exp\!\Big[\frac r2\big(a^{\dagger2}-a^2\big)\Big]\ket0.
 \label{eq:gaussian-input}
\end{equation}
Given $U$, $\eta$, $r$ and $\epsilon>0$, generate samples from a distribution $Q_U$ with $\E_{U\sim\mathrm{Haar}}\TV(P_U,Q_U)\le\epsilon$, where $P_U$ is the output distribution of this input under uniform loss with transmission $\eta$. Our main result considers all squeezed input states, $s=M$.
\end{problem}

An algorithm solves either problem efficiently if its running time per sample is polynomial in $M$ and $1/\epsilon$. Unless stated otherwise, $\eta$ and $r$ are fixed as the system grows, so the mean numbers of detected photons, $\eta n$ and $\eta s\sinh^2r$, grow in proportion to the number of sources. Running times count arithmetic operations, and Appendix~\ref{app:sampling} shows that polynomially many bits of precision suffice.

\section{Previous results}\label{sec:previous}

Most efficient classical algorithms for lossy boson sampling require the transmission to decrease as the system grows. At fixed total variation error, they apply when the mean number of detected photons grows at most as the square root of the number of sources. For single-photon inputs, the lossy state is then close to an efficiently simulable classical state~\cite{OszmaniecBrod2018,GarciaPatron2019}. For Gaussian inputs, the same scale appears for classical Gaussian approximations, whose bounds hold for every interferometer~\cite{Qi}, and in the loss analyses of tensor network methods~\cite{OhNP,OhMPS2026}. A recent Gaussian reset method, also valid for every interferometer, goes beyond this square root regime and allows a mean detected photon number of order $s^{2/3}$ for $s$ squeezed inputs~\cite{LimReset}.

Much less is known at inverse polynomial error. The Gaussian approximation and reset bounds then demand even stronger loss, and at the square root boundary the Gaussian tensor network analysis does not establish a cost polynomial in both system size and inverse error.

At a fixed transmission, interference truncation offers a different approach~\cite{Renema}. It approximates each output probability by keeping only interference among a limited number of photons, set by the transmission and the target error. The cost of evaluating this approximation grows as the system size raised to the truncation order, which is polynomial at fixed error but quasipolynomial at inverse polynomial error. Samples are then drawn with a Metropolis chain, for which no bound on the number of steps needed to reach a prescribed total variation error is given.

We obtained our results independently and became aware of the related preprint~\cite{GoJeong2026} after completing this work. That work establishes polynomial-time classical sampling for lossy Fock state boson sampling at any fixed survival probability in a dilute regime, with polynomial dependence on inverse error. In our notation, its full-output guarantee for $k$ survivors from $n$ input photons is proved under
\[
 M\ge C_0 n k^2,\qquad M\ge C_{\rm B}\frac{k^2}{\epsilon\delta},
\]
where $C_0,C_{\rm B}>0$ are universal constants, $\epsilon$ is the total variation error, and $\delta$ is the allowed failure probability over Haar random interferometers. At a fixed surviving fraction $k/n\in(0,1)$, the first bound already imposes $M=\Omega(n^3)$. The second bound makes the total collision probability small. Our result for Fock state boson sampling covers the full output distribution for every $M\ge n$, including $M=n$, with no additional condition on the relative scaling of mode and photon numbers. It includes all output collisions, even when they occur with substantial probability. Our Gaussian result further establishes efficient sampling with equal squeezing in all input modes.

Appendices~\ref{app:extensions} and \ref{app:hardness} compare our results with these simulation methods and with known hardness results.

\section{Main results}\label{sec:main-results}

\subsection{Efficient sampling at a fixed transmission}

Both problems of Sec.~\ref{sec:problem} admit efficient classical samplers at every fixed transmission. For single-photon inputs, the result is as follows.

\begin{theorem}[Efficient classical algorithm for lossy Fock state boson sampling]\label{thm:fock}
Fix a transmission $\eta$. For every $M\ge n$ and $0<\epsilon\le1/2$, there is a classical randomized algorithm for Problem~\ref{prob:fock} whose output distribution satisfies
\begin{equation}
 \E_{U\sim\mathrm{Haar}}\TV(P_U,Q_U)\le\epsilon.
 \label{eq:main-fock}
\end{equation}
Its worst case number of arithmetic operations per sample obeys
\begin{equation}
 T_{\mathrm F}(M,n,\epsilon)\le\operatorname{poly}_{\eta}(M,n,1/\epsilon).
 \label{eq:fcost}
\end{equation}
\end{theorem}

The polynomial's degree and constants may depend on the fixed transmission. The sampler works with any number of vacuum inputs and generates the full distribution of detected photon numbers. It therefore applies while the mean detected photon number $\eta n$ increases in direct proportion to the number of sources.

For Gaussian inputs squeezed in every mode, the same result holds at every finite squeezing.

\begin{theorem}[Efficient classical algorithm for lossy Gaussian boson sampling]\label{thm:gaussian}
Fix a transmission $\eta$ and a squeezing parameter $r\ge0$. For every $M\ge1$ and $0<\epsilon\le1/2$, there is a classical randomized algorithm for Problem~\ref{prob:gaussian} with $s=M$ whose output distribution satisfies
\begin{equation}
 \E_{U\sim\mathrm{Haar}}\TV(P_U,Q_U)\le\epsilon.
 \label{eq:allactive-guarantee}
\end{equation}
Its worst case number of arithmetic operations per sample obeys
\begin{equation}
 T_{\mathrm G}(M,\epsilon)\le\operatorname{poly}_{r,\eta}(M,1/\epsilon).
 \label{eq:gcost}
\end{equation}
\end{theorem}

At fixed $r>0$, the mean detected photon number $\eta M\sinh^2r$ grows linearly with $M$, as $\eta n$ does for single photons. The bound for a given $r$ also holds for every smaller squeezing parameter, so the squeezing may decrease with $M$, as in weak-squeezing scalings with $r\to0$. When only $s<M$ inputs are squeezed, Appendix~\ref{sec:vacuum} extends the result under the sufficient condition $\eta\cosh(2r)<1$, uniformly in $s$.

Both theorems bound the error on average over Haar random interferometers. A randomly drawn circuit therefore has a small error with high probability, although the algorithm cannot tell whether a given circuit does. By Markov's inequality, running the sampler at $\epsilon=\min\{1/2,M^{-(a+b)}\}$ for fixed $a,b>0$ gives an error at most $M^{-a}$ with probability at least $1-M^{-b}$ over the choice of $U$, still in polynomial time. Because the circuit is drawn once and then held fixed, all samples come from the same distribution $Q_U$ and carry the same error. The error of $T$ independent samples taken together is at most $T$ times this error. Running the sampler at an error smaller by a factor of $T$ therefore keeps polynomially many samples jointly accurate, still in polynomial time, as shown in Appendix~\ref{sec:model}. Section~\ref{sec:mechanism} outlines the proofs, which the appendices complete.

\subsection{Consequences for computational complexity}

Our results settle the regime of constant loss rate at the accuracy used in hardness arguments. For every fixed transmission, the samplers run in time polynomial in the system size and in $1/\epsilon$, which is the notion of efficient approximate sampling used for ideal boson sampling~\cite{AaronsonArkhipov}. In particular, the error may decrease as an inverse polynomial in the system size, measured by the photon number $n$ for single-photon inputs and by the mode number $M$ for Gaussian inputs. Choosing $\epsilon=n^{-a}$ or $\epsilon=M^{-a}$ with any fixed $a>0$ makes $1/\epsilon$ polynomial in the system size, so the running time remains polynomial. Earlier rigorous samplers needed the transmission to vanish as the system grows~\cite{OszmaniecBrod2018,GarciaPatron2019,Qi,LimReset}, and interference truncation becomes quasipolynomial at inverse polynomial error~\cite{Renema}. Our samplers extend efficient sampling to every constant loss rate. For Haar random interferometers, constant loss therefore leaves no asymptotic sampling advantage in either input family.

Hardness survives only when the loss rate vanishes quickly. Consider single-photon inputs with $M$ polynomial in $n$ and $(1-\eta)n=O(\log n)$. All $n$ photons then survive with probability $\eta^n\ge n^{-c}$ for some constant $c$, so keeping only the $n$-photon outcomes turns a sampler for the lossy distribution with sufficiently small inverse polynomial error into an ideal sampler at polynomial overhead~\cite{AaronsonBrod,GoDistinguishability}. The lossy problem is then as hard as ideal boson sampling under the same assumptions, and Appendix~\ref{app:hardness} gives the error and cost bounds of this reduction. The open range is thus confined to loss rates that vanish more slowly than $(\log n)/n$. Our results are consistent with this hardness. The correlation cutoff required by our samplers grows as the transmission approaches one, and at inverse polynomial error it becomes proportional to $n$ when only a logarithmic number of photons is lost on average. Because the cost grows exponentially with the cutoff, our bound on the running time is then no longer polynomial. For Gaussian inputs, a hardness comparison must further match the squeezing scale and interferometer ensemble of the relevant hardness result~\cite{GoLoss,Bouland}, and Appendix~\ref{app:hardness} treats these comparisons separately.

\subsection{Implications for quantum advantage experiments}

Hardness arguments for boson sampling take the interferometer from the Haar measure~\cite{AaronsonArkhipov,Hamilton}, and several experiments test their devices against this assumption. Jiuzhang 1.0 and 2.0 compared their measured transformation matrices with Haar random statistics~\cite{Zhong2020,Jiuzhang2021}, and a recent single-photon experiment implemented a unitary drawn from the Haar measure~\cite{Ascella2024}. Theorems~\ref{thm:fock} and \ref{thm:gaussian} apply to such experiments whenever their input states belong to the families treated there. In that case, adding photons and modes at a fixed transmission cannot establish an asymptotic sampling advantage, despite large photon counts, output collisions and multiphoton interference. Other large scale experiments use structured interferometers that are not Haar random, such as time domain loop networks~\cite{Borealis2022} and hybrid spatiotemporal circuits~\cite{Jiuzhang2026}, and our average case result does not directly apply to them.

Because the accuracy bound concerns the complete output distribution, it also holds after any processing applied in the same way to both outputs, such as recording only detector clicks or grouping output configurations into coarser classes. The probability that an output passes any given test therefore differs between the experiment and the classical sampler by at most the total variation distance. Our theorems nevertheless concern asymptotic scaling and do not by themselves show that any particular device can be spoofed. Spoofing a specific experiment would require an implementation that generates samples in a time comparable to that of the experiment, which polynomial bounds with possibly large degrees and constants do not ensure. It would also require an accuracy assessment for the actual circuit and physical noise model, because our bound holds on average over Haar random interferometers under uniform loss.

\begin{figure}[!tbp]
\centering
\includegraphics[width=\textwidth]{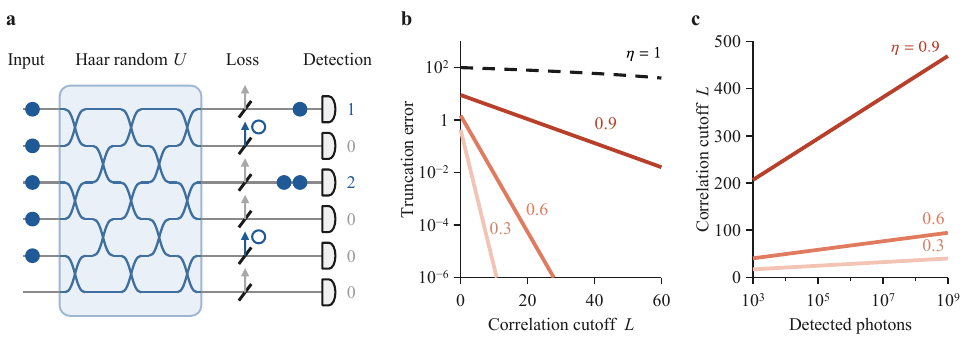}
\caption{\textbf{Loss suppresses correlations among many photons.}
\textbf{a}, Single photons enter a Haar random interferometer $U$, and each photon reaches the photon number resolving detectors with probability $\eta$.
\textbf{b}, Upper bound on the mean squared error, averaged over $U$, from keeping only correlations among at most $L$ photons, for $n=100$ single photons. Without loss ($\eta=1$) the error decreases only linearly in $L$, whereas loss makes it decay exponentially.
\textbf{c}, Cutoff $L$ needed for total variation error $1/n$. It grows only logarithmically with the number of detected photons.}
\label{fig:mechanism}
\end{figure}

\section{Sampling algorithms}\label{sec:mechanism}

Our samplers rest on one physical effect, the suppression of correlations among many photons by loss. Let $N$ be the total number of photons before loss. It equals $n$ for single-photon inputs and fluctuates over even values for Gaussian inputs. Each photon survives independently, so given $N$ the detected count $k$ is binomially distributed with $N$ trials and probability $\eta$. For each detected count, the samplers keep only correlation orders up to a cutoff $L$, evaluate the retained contributions efficiently, and convert them into samples by a rejection procedure with a bounded number of trials. The principle behind the method is that, with loss, the truncation error decays exponentially in $L$, so the required cutoff grows only logarithmically with the number of detected photons (Fig.~\ref{fig:mechanism}).

\subsection{Suppression of high-order correlations}

In Fig.~\ref{fig:mechanism}a, the loss acts in front of the detectors. Because uniform loss commutes with a passive interferometer, the same loss can be moved in front of the interferometer without changing the output distribution. For single-photon inputs, an event with $k$ surviving photons then amounts to an ideal experiment in which $k$ photons enter a uniformly random subset of the $n$ occupied input modes. Because the lost photons are not observed, the detection probabilities average these ideal experiments over all subsets, although each of them still shows full quantum interference.

For given $N$ and $k$, let $p_U(x)$ be the probability of the output configuration $x=(x_1,\ldots,x_M)$, which lists the photon number in each output mode, with $x_1+\cdots+x_M=k$. There are $D_{M,k}=\binom{M+k-1}{k}$ such configurations, and the rescaled probability $F_U(x)=D_{M,k}\,p_U(x)$ has the uniform average $D_{M,k}^{-1}\sum_xF_U(x)=\sum_xp_U(x)=1$. We expand it into orthogonal correlation orders,
\begin{equation}
 F_U=1+\sum_{h=1}^{k}\Pi_hF_U,\qquad W_h(k)=\E_U\|\Pi_hF_U\|_2^2 .
 \label{eq:main-decomp}
\end{equation}
Here $\|f\|_2^2=D_{M,k}^{-1}\sum_x|f(x)|^2$ is the uniform average of $|f|^2$, and $\Pi_h$ projects onto polynomials of degree $h$ in $x_1,\ldots,x_M$ that are orthogonal to all lower degrees under this average. The expectation of such a polynomial under $p_U$ depends only on the reduced state of $h$ photons, so $\Pi_hF_U$ describes the correlations among $h$ photons that remain after those among fewer photons are removed. For example, $\Pi_1F_U$ records how the mean photon number of each mode differs from the uniform value $k/M$, and $\Pi_2F_U$ records the pair correlations that these means do not explain. The weight $W_h(k)$ is its squared size averaged over Haar random interferometers.

For single-photon inputs, a correlation of order $h$ involves $h$ particular photons. These photons all survive in only a fraction $\binom{k}{h}/\binom{n}{h}$ of the possible $k$-photon subsets, and averaging over the subsets bounds $W_h(k)$ by this fraction. Its binomial average gives
\begin{equation}
 \E_kW_h(k)\le\eta^h,\qquad
 \E_{U,k}\|F_U-F_U^{\le L}\|_2^2\le\frac{\eta^{L+1}}{1-\eta},
 \label{eq:main-geometric}
\end{equation}
where $F_U^{\le L}$ keeps the orders $h\le L$. The truncation error thus decays exponentially in $L$ (Fig.~\ref{fig:mechanism}b). 

Gaussian inputs need one more step, because the number $N$ of photons before loss now fluctuates and the decay of the weights depends on the sector $(N,k)$. The weights again decay exponentially with $h$ when the photon density is bounded, $N\le cM$ for a fixed $c>\sinh^2r$, and the surviving fraction stays away from one, $k\le\vartheta N$ for a fixed $\vartheta$ with $\eta<\vartheta<1$. Sectors outside this range are rare when $N$ is large, because $N$ concentrates near its mean $M\sinh^2r$ and $k/N$ near $\eta$. When $N$ is small, all photons may survive with appreciable probability, but such sectors contain few photons and are sampled directly by computing their probabilities. Appendix~\ref{app:correlations} proves the bounds for both input families.

The truncated function $F_U^{\le L}$ need not be a probability distribution, but by the Cauchy--Schwarz inequality its mean absolute error is at most the square root of the bound in Eq.~\eqref{eq:main-geometric}. At fixed optical parameters, retaining orders up to
\begin{equation}
 L=O\!\left(1+\log(1/\epsilon)\right)
 \label{eq:main-cutoff}
\end{equation}
therefore keeps the truncation error below a fixed fraction of $\epsilon$, as shown in Appendix~\ref{app:sampling}. Inverse polynomial error requires only a logarithmic cutoff even when the number of detected photons grows linearly with size (Fig.~\ref{fig:mechanism}c).

\subsection{Efficient evaluation of low-order correlations}

To evaluate the retained orders, we write $p_U(x)$ as the overlap between the lossy input state and the projector onto $x$ carried back through the interferometer. Its component of order $h$ involves sums over $h$-photon configurations of the input and output modes, weighted by squared permanents of $h\times h$ submatrices of $U$ for single-photon inputs and by analogous matrix functions for Gaussian inputs. We call these configurations auxiliary, to distinguish them from the $k$-photon output configuration $x$ whose probability is being evaluated. Explicitly summing them at orders up to $L$ would cost $M^{O(L)}$, giving a quasipolynomial bound when $L$ grows logarithmically with size.

Our construction avoids this cost through two techniques. First, we project the factors on both the input and the output side onto the same correlation order before estimating their overlap. Since the projection is orthogonal, this leaves the contribution of that order unchanged, and the cancellations among lower orders are carried out exactly rather than through the random estimate. The estimation error is then controlled by the order weight $W_h(k)$. Second, every mode is assigned at random to one of $h$ groups. A set of $h$ distinct modes lands in distinct groups with probability $h!/h^h\ge \mathrm e^{-h}$, and the contributions of all such sets are summed at once by determinant generating functions. Reweighting by the inverse probability gives an unbiased estimate whose variance grows only exponentially in $h$. For single-photon inputs, the input factor also sums over which $h$ of the $n$ occupied input modes carry the photons, so these modes are grouped in the same way, independently of the output modes.

Auxiliary configurations with several photons in one mode need additional sums over their multiplicities, so we include them only up to a separate cutoff, chosen to keep the resulting bias below $\epsilon/16$. When the number of modes is at most a constant multiple of $L+\log(1/\epsilon)$, all such configurations can be included at polynomial cost. Otherwise they are rare under the uniform average, and a cutoff of order $L+\log(1/\epsilon)$ suffices. The estimator variance is at most exponential in $L$, and one evaluation costs $\operatorname{poly}(M,N,L)\exp[O(L+\log(1/\epsilon))]$ at fixed optical parameters. Together with Eq.~\eqref{eq:main-cutoff}, these bounds give polynomial dependence on system size and inverse error. Appendix~\ref{app:evaluation} develops the projection identities, determinant calculations and auxiliary cutoff bounds.

\subsection{Rejection sampling in polynomial time}

The algorithm first draws the source photon number $N$ and then the surviving count $k$, so that only the configuration within one photon sector remains to be sampled. The evaluation above yields, for any single output configuration $x$, an estimate of the truncated rescaled probability $F_U^{\le L}(x)$ with controlled bias and variance. It does not yield the distribution itself, because the number of configurations grows exponentially with the photon number. Rejection sampling turns these pointwise estimates into samples. Writing $F_U(x)=p_U(x)/u(x)$ with $u$ the uniform distribution on $k$-photon configurations, a uniformly proposed $x$ that is accepted with probability $F_U(x)/A$, for a constant $A\ge\max_xF_U(x)$, is distributed exactly according to $p_U$. Each trial is then accepted with probability $1/A$, so the expected number of trials is $A$.

This standard scheme is not efficient by itself, because $\max_xF_U(x)$ can be exponentially large in the photon number. We therefore clip the acceptance weight at a level $A$ of order $1/\epsilon$, which underweights the configurations with $F_U(x)>A$. Their total probability is at most $\|F_U\|_2^2/A$, since $F_U\le F_U^2/A$ on that set, and the second moment $\E_U\|F_U\|_2^2$ is bounded by a constant, $1/(1-\eta)$ for single-photon inputs. Clipping therefore changes the sampled distribution by an amount of order $1/A$ in total variation, on average over $U$, while each trial is still accepted with probability close to $1/A$. This bound on the second moment makes uniform proposals efficient.

In place of $F_U(x)$, the algorithm uses the mean of $T$ independent estimates of $F_U^{\le L}(x)$, clipped to $[0,A]$. Since the acceptance probability is linear in this weight, averaging over the internal randomness reproduces the intended acceptance probability up to an error set by the estimator variance. This variance is at most exponential in $L$, and since $L$ grows only logarithmically with $1/\epsilon$ by Eq.~\eqref{eq:main-cutoff}, it is polynomial in $1/\epsilon$. Fresh estimates are drawn for every proposal. The number of trials is capped at $J$ of order $A/\epsilon$, and a fixed configuration is output if all trials fail. The running time is therefore bounded for every circuit by $JT$ evaluations, which is polynomial in the system size and $1/\epsilon$.

A few photon sectors are handled separately. For Gaussian inputs, the source count is restricted to a finite range with a bounded omitted probability, and sectors with a small photon number before loss are sampled by sequential conditional probabilities followed by photon deletion. Other exceptional sectors are assigned fixed output configurations.

Appendix~\ref{app:sampling} bounds the total error, whose contributions come from truncating the correlation orders, omitting auxiliary configurations, the estimator noise, clipping at $A$, capping the number of trials at $J$, the exceptional photon sectors and finite numerical precision. With $L$, $T$, $A$ and $J$ chosen as described above, each contribution is a fixed fraction of $\epsilon$, so that $\E_U\TV(Q_U,P_U)\le\epsilon$ in time polynomial in the system size and $1/\epsilon$. In particular, the error can be made inverse polynomial in the system size while the running time stays polynomial.

\section{Discussion}\label{sec:discussion}
In experiments, a fixed transmission is the most favorable case that can realistically be reached. Each source, interferometer component and detector loses a fraction of the light that depends on the quality of that component rather than on the size of the system. The transmission approaching one that the known hardness results require thus appears achievable only in theory. A higher fixed transmission raises the degree of our polynomial bound but keeps the scaling polynomial, so reducing loss alone cannot produce an asymptotic advantage for these interferometers and inputs.

Atomic boson samplers illustrate this point. They lose about five percent of the atoms, independently of the evolution time, under approximately Haar random dynamics~\cite{Atomic2024}, and therefore operate at a fixed transmission far closer to one than photonic devices. As noted in Sec.~\ref{sec:problem}, our conclusion still applies to them asymptotically, but with a much higher polynomial degree, and the mean number of lost particles stays small up to larger sizes.

An asymptotic advantage therefore requires changes beyond the scope of our theorems, for example structured interferometers, other input states, or additional elements such as optical nonlinearities~\cite{Clavina2026}. Whether any of these changes makes sampling at a fixed transmission classically hard remains open.

\begin{acknowledgments}
The author obtained the main theoretical results. GPT-6 and Fable-5.1 served as assistants under the author's direction, helping to explore and sharpen mathematical arguments, to check calculations and to edit the text. All final proofs were verified independently by the author, who bears full responsibility for the results and conclusions. This work was supported by the Institute of Information \& Communications Technology Planning \& Evaluation (IITP) grant funded by the Korean government (Ministry of Science and ICT (MSIT)) (No. RS-2024-00437284, No. RS-2025-02283189), and National Research Foundation of Korea (NRF) grant funded by the Korean government (MSIT) (No. RS-2024-00431768).
\end{acknowledgments}

\appendix
\numberwithin{equation}{section}
\section*{Appendices}
The appendices give the precise sampling statements, the correlation bounds and their evaluation, and the complete sampling error analysis. The extension to Gaussian inputs containing vacuum follows after the sampling construction. The final appendices discuss the scope of the results and comparisons with other simulation methods and hardness results.

\section{Definition of the lossy sampling task}\label{sec:model}
This appendix fixes the conventions behind Theorems~\ref{thm:fock} and \ref{thm:gaussian} beyond those of Sec.~\ref{sec:problem}. We write $\E_U$ for the average over the normalized Haar measure $\mu_{\mathrm H}$ on $\mathrm U(M)$, so that the error in the theorems is
\begin{equation}
 \E_U\TV(P_U,Q_U)
 =\int_{\mathrm U(M)}\TV(P_U,Q_U)\,d\mu_{\mathrm H}(U).
 \label{eq:haar-average}
\end{equation}
The two output distributions are thus compared at each fixed $U$ before the error is averaged over circuits, and the accuracy can vary between interferometers. Running times count arithmetic operations on the entries of $U$ and on the optical parameters, and Sec.~\ref{app:precision} extends the bounds to bit complexity.

\subsection{Input states and output distributions}
We write $P^{\mathrm F}_{U,n,\eta}$ and $P^{\mathrm G}_{U,s,r,\eta}$ for the lossy output distributions of the inputs in Problems~\ref{prob:fock} and \ref{prob:gaussian}, with $n\le M$ single photons and $s\le M$ squeezed inputs. Theorems~\ref{thm:fock} and \ref{thm:gaussian} bound $\E_U\TV(Q_U,P^{\mathrm F}_{U,n,\eta})$ and $\E_U\TV(Q_U,P^{\mathrm G}_{U,M,r,\eta})$, with worst case running times $T_{\mathrm F}$ and $T_{\mathrm G}$ per sample. The degree and constants of $\operatorname{poly}$ depend on the fixed optical parameters and can diverge as $\eta\to1$. Appendix~\ref{app:correlations} decomposes both distributions into sectors with fixed photon numbers before and after loss.

For Gaussian inputs, Theorem~\ref{thm:gaussian} assumes that every input mode is squeezed, $s=M$, with the same squeezing parameter $r\ge0$ and no coherent displacement. Known squeezing phases may differ between modes, because their phase rotations, chosen independently of the interferometer, can be absorbed into the supplied matrix $U$ without changing its Haar distribution. General mixed Gaussian inputs and unequal nonzero squeezing magnitudes are not covered.

\subsection{From average error to individual circuits}
Theorems~\ref{thm:fock} and \ref{thm:gaussian} bound the error averaged over Haar interferometers. For any fixed $a,b>0$, setting $\epsilon=\min\{1/2,M^{-(a+b)}\}$ and applying Markov's inequality gives
\begin{equation}
 \Pr_U[\TV(P_U,Q_U)>M^{-a}]\le M^{-b}.
 \label{eq:typical-accuracy}
\end{equation}
The probability refers to the initial choice of interferometer, and the algorithm does not identify the exceptional circuits.

For $T$ independent samples from a fixed circuit, let $\delta_U=\TV(P_U,Q_U)$. Then
\[
 \TV(P_U^{\otimes T},Q_U^{\otimes T})\le T\delta_U.
\]
Running the sampler at $\epsilon=\min\{1/2,M^{-(a+b)}/T\}$ therefore gives a joint error at most $M^{-a}$ for a fraction at least $1-M^{-b}$ of circuits, and $1/\epsilon=\max\{2,TM^{a+b}\}$ remains polynomial in $M$ as long as $T$ is polynomial in $M$.

\section{Correlation bounds under loss}\label{app:correlations}
We first decompose the photon counts and define the correlation orders of the output probabilities. We then state and prove a bound in terms of reduced state purity before deriving the geometric decay for single-photon and Gaussian inputs. The remaining subsections give the normalization calculations used in the Gaussian case.

\subsection{Bounding correlations by reduced state purity}\label{sec:purity}
\subsubsection{Photon number decomposition}\label{sec:photon-number}
For single-photon inputs, the photon number before loss is fixed at $N=n$. For Gaussian boson sampling, let $s$ be the number of equally squeezed modes, with the other $M-s$ input modes in vacuum. The main result has $s=M$, while the extension in Appendix~\ref{sec:vacuum} allows $s<M$. In the Fock basis, the input in Eq.~\eqref{eq:gaussian-input} reads
\begin{equation}
 \ket{\Psi_{\mathrm G}}=(\sech r)^{s/2}\exp\Big(\frac{\tanh r}{2}\sum_{i=1}^{s}a_i^{\dagger2}\Big)\ket0^{\otimes M},
\end{equation}
a superposition of states with different total photon numbers. Before loss, only even values $N=2n$ occur, with $n=0,1,\ldots$ counting photon pairs. The photon number distribution is
\begin{equation}
 \Pr(N=2n)
 =\frac{\sech^s r\,\tanh^{2n}r}{2^n n!}
 \prod_{\ell=0}^{n-1}(s+2\ell).\label{eq:countlaw}
\end{equation}
Empty products are understood to equal one.
The normalized component of the input state in the sector with total photon number $N=2n$ is
\begin{equation}
 \ket{\Psi_{s,2n}}
 =\frac{(A_s^\dagger)^n\ket0}
 {\sqrt{2^{-n}n!\prod_{\ell=0}^{n-1}(s+2\ell)}},
 \qquad A_s^\dagger=\frac12\sum_{i=1}^s a_i^{\dagger2}.\label{eq:pairsector}
\end{equation}

For both input families, each of the $N$ photons survives independently with probability $\eta$, so the detected photon number $k=\sum_ix_i$ obeys $k\mid N\sim\operatorname{Bin}(N,\eta)$. As in Sec.~\ref{sec:problem}, the loss channel acts immediately before the detectors. Let $z=(z_1,\ldots,z_M)$ be the configuration of photon counts at the output of the interferometer before this loss channel, with $\sum_i z_i=N$, and let $x=(x_1,\ldots,x_M)$ be the detected configuration after loss, with $\sum_i x_i=k$. Both configurations refer to the same output modes. For a given configuration $z$, the loss channel acts independently on each mode. Before conditioning on the total number of survivors, the probability of obtaining $x$ is
\[
 \Pr(x\mid z)
 =\left[\prod_i\binom{z_i}{x_i}\right]\eta^k(1-\eta)^{N-k},
 \qquad k=\sum_i x_i.
\]
The probability that exactly $k$ photons survive is $\binom Nk\eta^k(1-\eta)^{N-k}$ for every configuration $z$ with total photon number $N$. Conditioning on exactly $k$ survivors gives the probability of obtaining $x$ from $z$ in this sector,
\begin{equation}
 K_{N\to k}(x\mid z)
 =\frac{\Pr(x\mid z)}{\binom Nk\eta^k(1-\eta)^{N-k}}
 =\frac{\prod_i\binom{z_i}{x_i}}{\binom Nk},
 \quad x\le z,\quad |x|=k.\label{eq:deletion}
\end{equation}
Here $x\le z$ means $x_i\le z_i$ for every output mode, since loss removes photons within each output mode. Equality is allowed in a mode where no photon is lost. We write $|x|=\sum_i x_i$. The probability is zero outside these constraints. The factor $\eta^k(1-\eta)^{N-k}$ cancels, so $K_{N\to k}$ does not depend on the transmission and every set of $k$ surviving photons is equally likely.

The interferometer determines the probabilities of the configurations $z$. Let $p_U^{(N)}(z)$ denote the output distribution without loss in the $N$-photon sector. Since the probability of $k$ survivors is the same for every $z$, the detected distribution conditioned on $N$ and $k$ is
\[
 p_U^{(N,k)}(x)
 =\sum_{z:\,|z|=N}p_U^{(N)}(z)\,K_{N\to k}(x\mid z).
\]
Thus $K_{N\to k}$ describes only the loss transition for a given $z$, while the dependence on $U$ enters through $p_U^{(N)}$. This is a decomposition of the measured photon count probabilities for the original circuit.

Within a fixed $N$-photon sector, uniform loss conditioned on $k$ survivors gives the normalized $k$-particle reduced state of the state leaving the interferometer. Equation~\eqref{eq:deletion} describes its action on photon count probabilities, including collisions~\cite{Ivan}. The adjoint loss channel maps each photon counting effect to an operator diagonal in photon number, so coherence between states with different input photon numbers $N$ does not contribute to the measured probabilities. Sampling Eq.~\eqref{eq:countlaw} and then $k$ therefore gives an exact classical decomposition of the complete output distribution.

\subsubsection{Correlation decomposition of output probabilities}
In each photon number sector, the loss acting after the interferometer can be moved to the input, as we now show. Let $\rho_N$ be the normalized input state in a fixed $N$-photon sector. The Hilbert space of $m$ identical bosons is $\Sym^m(\mathbb C^M)$, the subspace of $(\mathbb C^M)^{\otimes m}$ invariant under particle permutations. The interferometer acts on this space as
\[
 \mathcal U_m=\Sym^m(U)
 =\left.U^{\otimes m}\right|_{\Sym^m(\mathbb C^M)}.
\]
Conditioned on $k$ surviving photons, the normalized state after the output loss is
\begin{equation}
 \rho^{\mathrm{out}}_{U,N,k} =\Tr_{N-k}\!\left(\mathcal U_N\rho_N\mathcal U_N^\dagger\right) =\mathcal U_k\sigma_{N,k}\mathcal U_k^\dagger, \qquad \sigma_{N,k}=\Tr_{N-k}\rho_N.
 \label{eq:particle-covariance}
\end{equation}
Here $\Tr_{N-k}$ traces out $N-k$ particles in the first-quantized description, leaving a reduced state on all $M$ modes. Since $U^{\otimes N}=U^{\otimes k}\otimes U^{\otimes(N-k)}$, the unitary acting on the traced particles drops out of the partial trace, leaving $\mathcal U_k$ on the remaining particles. Thus $\sigma_{N,k}$ is a reduced state of the original input and is independent of $U$. This is the operator form of the deletion in Eq.~\eqref{eq:deletion}, with the reduction acting on the input. For a fixed $(N,k)$ sector, we abbreviate $\sigma_{N,k}$ as $\sigma_k$. More generally, the bounds below apply to any normalized $k$-boson state $\sigma_k$ held fixed as $U$ varies.

The possible output configurations, their number, and the uniform reference probability are
\begin{equation}
 \Omega_{M,k}=\{x\in\mathbb N^M:|x|=k\},\qquad D_{M,k}=\binom{M+k-1}{k},\qquad u_{M,k}=D_{M,k}^{-1}.
\end{equation}
Here the entries of $x$ are nonnegative integers, and each configuration labels a Fock basis state $\ket{x}$ of the $k$-photon Hilbert space. Thus $D_{M,k}$ is also the dimension of this space. With $p_U(x)=\bra{x}\mathcal U_k\sigma_k\mathcal U_k^\dagger\ket{x}$ denoting the output probability, we define the rescaled output probability
\begin{equation}
 F_U(x)=\frac{p_U(x)}{u_{M,k}}=D_{M,k}\,p_U(x).
\end{equation}
This rescaling makes the uniform distribution the constant function $F_U=1$. For every $U$, $F_U\ge0$ and $\E_uF_U=1$, where $\E_u$ denotes the uniform average over $\Omega_{M,k}$. The superscript in $F_U^{(m)}$ specifies the photon number $m$.

Under interferometer conjugation $X\mapsto\mathcal U_kX\mathcal U_k^\dagger$, the space of linear operators on the $k$-photon Hilbert space decomposes into orthogonal correlation sectors $h=0,\ldots,k$. These are irreducible sectors, the smallest nonzero operator subspaces preserved by every interferometer, and each occurs once. Orthogonality is with respect to $\Tr(X^\dagger Y)$.

Operators of order at most $h$ correspond to operators on the $h$-photon space, a space of dimension $D_{M,h}^2$, so the full order $h$ sector has dimension
\begin{equation}
 R_{M,h}=D_{M,h}^2-D_{M,h-1}^2,\label{eq:irrepdim}
\end{equation}
Restricting to operators diagonal in the Fock basis, which photon counting probes, replaces $D_{M,h}^2$ by $D_{M,h}$ and gives the dimension of the Fock-diagonal part,
\begin{equation}
 Z_{M,h}=D_{M,h}-D_{M,h-1}
 =\binom{M+h-2}{h},\label{eq:diagdim}
\end{equation}
where $D_{M,-1}=0$~\cite{Arienzo}.

On output configurations, these sectors correspond to multivariate Hahn polynomials, which have degree $h$ in the occupations and are orthogonal to all lower degrees under $u_{M,k}$. Let $\Pi_h$ denote the projection onto order $h$ in the space of functions on $\Omega_{M,k}$, with inner product and norm
\[
 \langle f,g\rangle_u
 =\frac{1}{D_{M,k}}\sum_{x\in\Omega_{M,k}}f(x)^*g(x),
 \qquad \|f\|_2^2=\langle f,f\rangle_u.
\]
The constant component is $\Pi_0F_U=1$, so
\begin{equation}
 F_U=1+\sum_{h=1}^{k}\Pi_hF_U,\qquad W_h(k)=\E_U\|\Pi_hF_U\|_2^2,
 \label{eq:app-decomp}
\end{equation}
where $W_h(k)$ is the mean squared contribution of order $h$. For an integer cutoff $L\ge0$, define
\[
 F_U^{\le L}:=1+\sum_{h=1}^{\min(L,k)}\Pi_hF_U.
\]
The superscript $\le L$ specifies the cutoff in correlation order, while the detected photon number $k$ can still be much larger than $L$. Orthogonality gives
\[
 \|F_U-F_U^{\le L}\|_2^2
 =\sum_{h=L+1}^{k}\|\Pi_hF_U\|_2^2,
\]
with an empty sum equal to zero. The projections depend only on $M$ and $k$ and are fixed before averaging over $U$.

\subsubsection{A bound from reduced state purity}
The weight $W_h(k)$ in Eq.~\eqref{eq:app-decomp} can be bounded in terms of the normalized $h$-particle reduced state $\gamma_h=\Tr_{k-h}\sigma_k$ and its purity $\tau_h=\Tr\gamma_h^2$. For $M\ge2$ and $1\le h\le k$,
\begin{equation}
 W_h(k)\le\binom kh \frac{M+h-1}{M+2h-1} \frac{(M+2h-1)!(M+k-1)!} {(M+h-1)!(M+k+h-1)!}\,\tau_h \le\binom kh\tau_h.\label{eq:puritybound}
\end{equation}
The only quantity in this bound that depends on the state is $\tau_h$, so the correlation estimate reduces to a calculation on the reduced input state. The bound holds for any normalized bosonic state $\sigma_k$, and adding vacuum modes leaves $\tau_h$ unchanged. The proof follows in Sec.~\ref{app:representation}.

\subsection{Proof of the purity bound}\label{app:representation}
The proof has two steps. We first evaluate the Haar average of a single correlation sector and then determine how particle reduction rescales each correlation order. Combining the two results for the reduced state $\gamma_h$ gives the bound in Eq.~\eqref{eq:puritybound}.

\subsubsection{Haar average of a correlation sector}
Let $\rho$ be a normalized $N$-photon state and $\mathcal P_h^{\rm op}$ the orthogonal projection onto the order $h$ operator sector defined above. Its correlation components are
\[
 \rho^{[h]}=\mathcal P_h^{\rm op}\rho,
 \qquad \rho=\sum_{h=0}^{N}\rho^{[h]},
 \qquad \rho^{[0]}=\frac{I}{D_{M,N}}.
\]
The index $h$ labels the correlation order of $\rho^{[h]}$, which acts on the full $N$-photon space. For $h>0$, these components are traceless Hermitian operators, so each nonzero component has both positive and negative eigenvalues and is not a state by itself. The order $h$ contribution to $F_U^{(N)}$ satisfies
\[
 \|\Pi_hF_U^{(N)}\|_2^2
 =D_{M,N}\sum_{|x|=N}
 \left|\bra{x}\mathcal U_N\rho^{[h]}\mathcal U_N^\dagger\ket{x}\right|^2.
\]
Haar orthogonality gives the same mean squared overlap with each normalized operator direction in the irreducible order $h$ sector. Photon counting probes its $Z_{M,h}$ diagonal directions out of $R_{M,h}$ directions in total, yielding
\begin{equation}
 \E_U\|\Pi_h F_U^{(N)}\|_2^2
 =\frac{D_{M,N}Z_{M,h}}{R_{M,h}}
    \|\rho^{[h]}\|_{\HS}^2.\label{eq:readoutnorm}
\end{equation}
Here $\|X\|_{\HS}^2=\Tr(X^\dagger X)$ is the squared Hilbert--Schmidt norm. The factor $D_{M,N}$ comes from the normalization of $F_U^{(N)}$, and the sector dimensions are given in Eqs.~\eqref{eq:irrepdim} and \eqref{eq:diagdim}~\cite{Arienzo}.

\subsubsection{Effect of particle reduction}\label{sec:reduction}
By Eq.~\eqref{eq:particle-covariance}, applying $\mathcal U_N$ before reduction gives the same state as applying $\mathcal U_k$ afterwards. Because each correlation sector is irreducible and occurs once, reduction acts separately on each order and rescales all operator directions within that order equally. For $0\le h\le k\le N$, let $s_h(M,N,k)\ge0$ denote the resulting norm factor for $F_U$,
\[
 \|\Pi_hF_U^{(k)}\|_2^2
 =s_h^2(M,N,k)\,\|\Pi_hF_U^{(N)}\|_2^2,
\]
where each norm uses the uniform reference distribution at its own photon number. This identity holds for each $U$. On photon count probabilities, the reduction is the deletion map in Eq.~\eqref{eq:deletion}.

Since $s_h(M,N,k)$ is the same for every operator of order $h$, it suffices to follow one nonzero operator of that order. At photon number $n\ge h$, choose
\begin{equation}
 E_{n,h}=(a_1^\dagger)^h a_M^h\big|_{\Sym^n\mathbb C^M}.
\end{equation}
Here $a_i$ annihilates a photon in mode $i$, so $E_{n,h}$ transfers $h$ photons from mode $M$ to mode $1$. Its expectation after reduction obeys
\[
 \Tr\!\left[E_{k,h}\Tr_{N-k}\rho\right]
 =\frac{\binom{k}{h}}{\binom{N}{h}}\Tr(E_{N,h}\rho).
\]
The ratio is the probability that a specified set of $h$ particles is included in a uniformly chosen subset of $k$ out of $N$ particles. To obtain $s_h$, we must also account for the different operator norms at photon numbers $N$ and $k$.

The squared Fock basis matrix elements of $E_{n,h}$ sum to
\[
 \|E_{n,h}\|_{\HS}^2
 =(h!)^2\sum_{|x|=n}\binom{x_1+h}{h}\binom{x_M}{h}
 =(h!)^2\binom{M+n+h-1}{n-h},
\]
where the last equality follows from binomial convolution. Dividing the squared norm by $D_{M,n}$ gives
\begin{equation}
 \frac{\|E_{n,h}\|_{\HS}^2}{D_{M,n}}
 =\frac{(h!)^2 n!(M+n+h-1)!(M-1)!}
 {(n-h)!(M+n-1)!(M+2h-1)!}.\label{eq:highestnorm}
\end{equation}
Combining the expectation ratio with these norms in the normalization relative to the uniform reference yields
\begin{equation}
 s_h^2(M,N,k) =\left(\frac{\binom{k}{h}}{\binom{N}{h}}\right)^2 \frac{\|E_{N,h}\|_{\HS}^2/D_{M,N}} {\|E_{k,h}\|_{\HS}^2/D_{M,k}} =\frac{\binom{k}{h}}{\binom{N}{h}} \frac{(M+N+h-1)!(M+k-1)!} {(M+N-1)!(M+k+h-1)!}.\label{eq:generalsingular}
\end{equation}
The factorial ratio accounts for the change in normalization between photon sectors.

\subsubsection{Combining the two identities}
Return to the $k$-photon state $\sigma_k$ and its reduction $\gamma_h$. Setting the initial and final photon numbers in Eq.~\eqref{eq:generalsingular} to $k$ and $h$ gives
\begin{equation}
 s_h^2(M,k,h)=\frac{1}{\binom{k}{h}}
 \frac{(M+k+h-1)!(M+h-1)!}
 {(M+k-1)!(M+2h-1)!}.\label{eq:losssingular}
\end{equation}
On the $h$-photon space, write $\gamma_h^{[h]}=\mathcal P_h^{\rm op}\gamma_h$ for the order $h$ component of the reduced state. Applying Eq.~\eqref{eq:readoutnorm} to $\gamma_h$ yields
\begin{equation}
 s_h^2(M,k,h)W_h(k) =\frac{D_{M,h}Z_{M,h}}{R_{M,h}} \|\gamma_h^{[h]}\|_{\HS}^2 \le\frac{D_{M,h}Z_{M,h}}{R_{M,h}}\Tr\gamma_h^2 =\frac{M+h-1}{M+2h-1}\tau_h.
\end{equation}
The inequality follows because orthogonal projection cannot increase the squared norm, and $\|\gamma_h\|_{\HS}^2=\Tr\gamma_h^2$. Dividing by Eq.~\eqref{eq:losssingular} proves the first inequality in Eq.~\eqref{eq:puritybound}. The second follows from $(M+h-1)/(M+2h-1)\le1$ and the factorial ratio $\prod_{\ell=0}^{h-1}(M+h+\ell)/(M+k+\ell)\le1$ for $h\le k$.

\subsection{Fock state boson sampling}\label{sec:fproof}
For one photon in each of $n$ distinct input modes, the normalized $h$ particle reduction is
\begin{equation}
 \gamma_h=\binom nh^{-1}
 \sum_{\substack{S\subseteq[n]\\|S|=h}}
 \ket{1_S}\bra{1_S}.
\end{equation}
The components are orthogonal, hence
\begin{equation}
 \tau_h=\binom nh^{-1}.\label{eq:fockpurity}
\end{equation}
Conditioned on $k\sim\operatorname{Bin}(n,\eta)$, the input after loss is $\gamma_k$. For $h\le k$, its $h$ particle reduction is again $\gamma_h$, and Eq.~\eqref{eq:puritybound} gives
\begin{equation}
 W_h(k)\le\frac{\binom kh}{\binom nh}.
\end{equation}
For $h>k$, there is no order $h$ component, so $W_h(k)=0$ and the bound still holds with $\binom kh=0$. The binomial factorial moment identity then gives
\begin{equation}
 \E_kW_h(k)\le\eta^h.\label{eq:fockgeometric}
\end{equation}
Averaging over Haar random $U$ and the binomial count $k$, orthogonality and Eq.~\eqref{eq:fockgeometric} bound the squared truncation error and, after adding the constant component $\|\Pi_0F_U\|_2^2=1$, the second moment,
\begin{align}
 \E_{U,k}\|F_U-F_U^{\le L}\|_2^2
 &=\sum_{h=L+1}^{n}\E_kW_h(k)\le\sum_{h=L+1}^{\infty}\eta^h
 =\frac{\eta^{L+1}}{1-\eta},\label{eq:focktail}\\
 \E_{U,k}\|F_U\|_2^2&\le\frac1{1-\eta}.\label{eq:focksecond}
\end{align}
Both bounds hold at every finite $M$, because they rest on the exact Haar average in Eq.~\eqref{eq:readoutnorm} rather than on an approximation for large $M$, and the average over $k$ includes the lossless event $k=n$.

\subsection{Gaussian boson sampling}\label{sec:gproof}
We now prove the bounds for Theorem~\ref{thm:gaussian}, in which every input mode is squeezed, $s=M$, for every fixed $r>0$ and $\eta$. We fix a photon number sector $N=2n$ before loss. This suffices, because the measured probabilities are a mixture over such sectors with the weights in Eq.~\eqref{eq:countlaw}, as shown in Sec.~\ref{sec:photon-number}. For the normalized $N$-photon input component $\ket{\Psi_{M,N}}$ in Eq.~\eqref{eq:pairsector}, let
\[
 \Gamma_{M,N,h}:=\E_U\|\Pi_hF_U^{(N)}\|_2^2.
\]
This is the weight $W_h(k)$ of Eq.~\eqref{eq:app-decomp} for the lossless case $k=N$. Odd orders vanish because the input sector is invariant under real orthogonal mode transformations, while for $h=2j$ with $0\le j\le n$,
\begin{equation}
 \Gamma_{M,2n,2j} =\frac{b_jb_{n-j}}{b_n} \prod_{v=0}^{j-1} \frac{M+2v-1}{M+2v} \times \frac{M+2n+2v}{M+2n+2v+1},\qquad b_j=4^{-j}\binom{2j}{j}.\label{eq:coreweights}
\end{equation}
This is the equal squeezing coefficient of Ref.~\cite{Dou}, rescaled to our normalization of $F_U$. Section~\ref{app:core} gives the normalization calculation.

Vandermonde's identity implies
\begin{equation}
 \binom{2n}{n}\ge\binom{2j}{j}\binom{2n-2j}{n-j},
\end{equation}
so the prefactor $b_jb_{n-j}/b_n$ is at most one. Every other factor in Eq.~\eqref{eq:coreweights} is also at most one, so
\begin{equation}
 0\le \Gamma_{M,N,h}\le1 \qquad(0\le h\le N).\label{eq:coreone}
\end{equation}
Without loss, each order up to $N$ can therefore carry a weight of order one, so the number of relevant orders can grow with $N$.

Reduction from $N$ to $k$ photons gives the exact weight
\begin{equation}
 W_h(k)=\Gamma_{M,N,h}
 \frac{\binom{k}{h}}{\binom{N}{h}}
 \frac{(M+N+h-1)!(M+k-1)!}
 {(M+N-1)!(M+k+h-1)!}.\label{eq:fullactiveweight}
\end{equation}
The factor multiplying $\Gamma_{M,N,h}$ is $\prod_{\ell=0}^{h-1}(k-\ell)(M+N+\ell)/[(N-\ell)(M+k+\ell)]$. Each factor of this product is at most its value at $\ell=0$, so
\begin{equation}
 W_h(k)\le\chi^h,
 \qquad \chi=\frac{k(M+N)}{N(M+k)}.
\end{equation}
On the count region $N\le cM$ and $k\le\vartheta N$, where $c$ is fixed and $\eta<\vartheta<1$,
\begin{equation}
 \chi\le\chi_*:=\frac{\vartheta(1+c)}{1+\vartheta c}<1.
\end{equation}
For the cutoff $L=2\ell$, orthogonality and the absence of odd orders bound the squared truncation error and, after adding the constant component, the second moment,
\begin{align}
 \E_U\|F_U-F_U^{\le2\ell}\|_2^2
 &\le\frac{\chi_*^{2\ell+2}}{1-\chi_*^2},\label{eq:alltail}\\
 \E_U\|F_U\|_2^2&\le\frac1{1-\chi_*^2}.\label{eq:allsecond}
\end{align}
For fixed $r$, the mean input photon number is $M\sinh^2r$, and Sec.~\ref{app:counts} treats the sectors outside this count region.

\subsection{Haar coefficients for equal squeezing}\label{app:core}
We take $M\ge2$ and follow the derivation in Ref.~\cite{Dou} to compute the coefficient in Eq.~\eqref{eq:coreweights}.

Under real orthogonal mode transformations, the $k$-photon space has the Fischer decomposition
\begin{equation}
 \Sym^k\mathbb C^M
 =\bigoplus_{q=0}^{\lfloor k/2\rfloor}
 (A_M^\dagger)^q\,\mathrm{Harm}_{k-2q},
 \label{eq:fischer}
\end{equation}
where $\mathrm{Harm}_j$ is the space of $j$-photon states annihilated by the pair annihilation operator $A_M=\frac12\sum_ia_i^2$. In the Bargmann representation, these states are the harmonic polynomials of degree $j$. On the $h$-photon space, an operator $X$ has correlation order exactly $h\ge1$ precisely when $\Tr_1 X=0$, where $\Tr_1$ traces out one particle. An operator invariant under all real orthogonal mode transformations commutes with their action, so by Schur's lemma it is a multiple of the identity on each summand of Eq.~\eqref{eq:fischer}. On the $h$-photon space, such operators therefore form a space of dimension $\lfloor h/2\rfloor+1$, and those of order at most $h-1$ form a space of dimension $\lfloor(h-1)/2\rfloor+1$. The invariant operators of order exactly $h$ thus form a one dimensional space for even $h$ and none exist for odd $h$. The state $\rho_{2n}=\ket{\Psi_{M,2n}}\bra{\Psi_{M,2n}}$ is orthogonally invariant, and so is each of its correlation components. Its odd components therefore vanish, and each even component lies in the one dimensional invariant space of its order.

For $j\ge1$, choose the order $2j$ operator $Y_j$ and average it over real orthogonal mode transformations,
\begin{equation}
 Y_j=\ket{2je_1}\bra{2je_2},\qquad
 \overline Y_j=\int_{O(M)}\Sym^{2j}(O)Y_j\Sym^{2j}(O)^\dagger\,dO.
\end{equation}
Here $\ket{2je_i}$ denotes the Fock state with $2j$ photons in mode $i$ and vacuum in every other mode. Since $\Tr_1 Y_j=0$, $Y_j$ lies entirely in order $2j$, and its average $\overline Y_j$ spans the invariant subspace. The squared Hilbert--Schmidt norm of $\overline Y_j$ is
\begin{equation}
 \beta_{M,j} =\int_{O(M)}O_{11}^{2j}O_{22}^{2j}\,dO =\frac{[(2j)!]^2}{4^j(j!)^2} \prod_{\ell=0}^{j-1} \frac{M+2j+2\ell-1} {(M+2\ell-1)(M+2\ell)(M+2j+2\ell)}.\label{eq:beta}
\end{equation}
To obtain this integral, condition on the first column $x$ of $O$. The second column is then uniform on the unit sphere orthogonal to $x$, and its relevant moment is
\[
 (1-x_2^2)^j\frac{(2j)!}
 {2^j j!\prod_{\ell=0}^{j-1}(M-1+2\ell)}.
\]
Expanding the power and averaging over $x$ with
\begin{equation}
 \E x_1^{2a}x_2^{2b}
 =\frac{(2a)!(2b)!}
 {2^{a+b}a!b!\prod_{\ell=0}^{a+b-1}(M+2\ell)}
\end{equation}
gives a finite series, which the Chu--Vandermonde identity sums to Eq.~\eqref{eq:beta}.

Write $\mathcal D(X)=\sum_i a_iXa_i^\dagger$ for unnormalized removal of one particle. On an $N$-photon state, normalized reduction to $k$ photons is $(k!/N!)\mathcal D^{N-k}$. Expanding $\rho_{2n}$ in the Fock basis and deleting $2n-2j$ photons gives
\begin{equation}
 \mu_{M,n,j} :={\langle Y_j,\mathcal D^{2n-2j}(\rho_{2n})\rangle}_{\HS} =\frac{(2n-2j)!n!(2j)!}{2^j(j!)^2(n-j)!} \frac{\prod_{\ell=0}^{n-j-1}(M+4j+2\ell)} {\prod_{\ell=0}^{n-1}(M+2\ell)}.\label{eq:mu}
\end{equation}
This follows from the Fock coefficients
\begin{equation}
 \langle2\bm q|\Psi_{M,2n}\rangle
 =\sqrt{\frac{n!}{2^n\prod_{\ell=0}^{n-1}(M+2\ell)}}
   \prod_i\frac{\sqrt{(2q_i)!}}{q_i!}
\end{equation}
and coefficient extraction in
\begin{equation}
 \left[\frac{(2j)!}{j!}(1-4z)^{-j-1/2}\right]^2
 (1-4z)^{-(M-2)/2}.
\end{equation}
The Hilbert--Schmidt adjoint $\mathcal D^*$ adds one particle, and after $2n-2j$ additions
\begin{equation}
 \|(\mathcal D^*)^{2n-2j}\overline Y_j\|_{\HS}^2
 =\frac{(2n-2j)!(M+2n+2j-1)!}{(M+4j-1)!}\beta_{M,j}.
\end{equation}
Since $(\mathcal D^*)^{2n-2j}\overline Y_j$ spans the invariant subspace of order $2j$ in the $2n$-photon space,
\begin{equation}
 \|\rho_{2n}^{[2j]}\|_{\HS}^2
 =\frac{(M+4j-1)!\,\mu_{M,n,j}^2}
 {(2n-2j)!(M+2n+2j-1)!\,\beta_{M,j}}.
\end{equation}
Multiplying by $D_{M,2n}/R_{M,2j}$ as in Eq.~\eqref{eq:readoutnorm} gives the coefficient
\begin{equation}
 \Lambda_{2j} =\left[\frac{(2j)!n!}{j!(n-j)!}\right]^2 \frac{(2n-2j)!}{(2n)!} \times\prod_{\ell=0}^{j-1} \frac{M+2n+2\ell} {(M+2\ell)^2(M+2n+2\ell+1)}.
\end{equation}
Multiplying by $Z_{M,2j}$, as Eq.~\eqref{eq:readoutnorm} requires, and separating the even and odd factors yields Eq.~\eqref{eq:coreweights}. The constant order has $\Gamma_{M,N,0}=1$.

\section{Evaluation of low-order correlations}\label{app:evaluation}
Appendix~\ref{app:correlations} shows that the output distribution can be approximated by truncating its correlation expansion at order $L$. To turn this approximation into a sampler, we must compute a weight for each proposed output. The task here is therefore to estimate $F_U^{\le L}(x)$ from the supplied interferometer $U$ and a proposed configuration $x$, at fixed photon numbers $N$ before loss and $k=|x|$ after loss.

We construct the estimate one correlation order at a time. First, particle reduction expresses each contribution as a sum over configurations of only $h\le\min\{L,k\}$ photons, which we call auxiliary configurations. We then group the mode labels at random so that the terms in this sum can be combined by generating functions. We define the resulting random estimate before proving its error bounds and giving its explicit evaluation for single-photon and Gaussian inputs. Finally, we choose the auxiliary cutoff and bound the total evaluation cost. In Appendix~\ref{app:sampling}, these estimates set the acceptance probabilities of the sampler.

\subsection{Constructing the random estimate}\label{sec:color}
\subsubsection{Direct evaluation and its limitations}
A direct reconstruction uses reduced photon count probabilities. Let $P_{N\to h}(y)$ be the probability of the $h$-photon configuration $y$ obtained by reducing the interferometer output in the fixed $N$-photon sector. The configuration $y$ is an auxiliary summation index used to evaluate the weight of the proposed $k$-photon output $x$. We suppress $U$ in this notation and write $F_U^{(h)}(y)=D_{M,h}P_{N\to h}(y)$. 

To separate the correlation orders, we use an operator that forgets all but $h$ of the $k$ photons and then refills the configuration. It acts on functions $f$ of $k$-photon configurations by $(\mathcal T_{k,h}f)(x)=\E f(x')$, where $x'$ is obtained from $x$ by keeping a uniformly chosen set of $h$ photons and then adding $k-h$ photons one at a time. If the current configuration is $z$, choose mode $i$ with probability $(z_i+1)/(M+|z|)$, increase $z_i$ by one, and update the probabilities before the next insertion. Under the uniform reference distributions, this rule is the reverse of uniform photon deletion, so reinsertion and deletion are adjoint. The resulting operator preserves $u_{M,k}$ and is self-adjoint. As shown in Sec.~\ref{sec:reduction}, deletion acts on each correlation order separately and rescales it by a single factor, and the same holds for reinsertion. Each order $j\le h$ is therefore an eigenspace of $\mathcal T_{k,h}$, and Eq.~\eqref{eq:generalsingular} gives its eigenvalue
\begin{equation}
 \lambda_{h,j}
 =\frac{\binom{h}{j}}{\binom{k}{j}}
 \frac{(M+k+j-1)!(M+h-1)!}{(M+k-1)!(M+h+j-1)!},
 \qquad j\le h.\label{eq:du}
\end{equation}
Orders above $h$ are annihilated, because only $h$ photons are retained. For this reconstruction take $0\le L\le k$, replacing a larger cutoff by $k$. A triangular system gives $F_U^{\le L}=\sum_{h\le L}c_h\mathcal T_{k,h}F_U$, where
\begin{equation}
 \sum_{h=j}^{L}c_h\lambda_{h,j}=1,
 \qquad 0\le j\le L.\label{eq:triangular}
\end{equation}
Since $\lambda_{h,h}\ne0$, these coefficients can be found by backward substitution. Composition of normalized particle reductions gives the direct evaluation formula
\begin{equation}
 (\mathcal T_{k,h}F_U)(x) =\frac{D_{M,h}}{\binom kh} \sum_{\substack{y\le x\\|y|=h}} \left[\prod_i\binom{x_i}{y_i}\right]P_{N\to h}(y).\label{eq:evaluation}
\end{equation}
Direct evaluation costs $(M+N+L)^{O(L)}$. Moreover, the coefficient
\[
 c_L=\binom{k}{L}
 \frac{(M+2L-1)!(M+k-1)!}{(M+L-1)!(M+k+L-1)!}
\]
can be large and amplify the errors of independent estimates of the summands. We therefore isolate each correlation order in both factors before estimating the sum.

\subsubsection{Two-sided projection}
We first rewrite the contribution at order $h$ so that the cancellations among lower orders are carried out before random estimation. This will allow the error to be controlled by the order weight $W_h(k)$ from Appendix~\ref{app:correlations}. Normalization fixes the component of order zero to $\Pi_0F_U=1$, so it requires no estimation. We therefore consider $1\le h\le k$.

Let $D_h=D_{M,h}$ and let $\mathcal H_h$ be the space of functions on $h$-photon configurations, with the uniform inner product used in Appendix~\ref{app:correlations}. In this space, $\Pi_h$ projects onto order $h$ and removes all lower orders. Its range has dimension $Z_{M,h}=D_h-D_{h-1}$, as in Eq.~\eqref{eq:diagdim}. For a proposed output $x$, the probability of obtaining $y$ by selecting $h$ photons uniformly without replacement is $\prod_i\binom{x_i}{y_i}/\binom kh$. Define the corresponding weight $w_x$ and its order $h$ part $v_x$, and write $\psi_h(U)$ for the order $h$ part of the reduced output $F_U^{(h)}$ defined above,
\begin{equation}
 w_x(y)=D_h\frac{\prod_i\binom{x_i}{y_i}}{\binom kh},\qquad
 v_x=\Pi_hw_x,\qquad
 \psi_h(U)=\Pi_hF_U^{(h)}.
 \label{eq:framesetup}
\end{equation}
Thus $v_x$ is determined by the proposed configuration, while $\psi_h(U)$ is determined by the optical output. With $\lambda_h=\lambda_{h,h}$ from Eq.~\eqref{eq:du}, particle reduction gives
\begin{equation}
 f_h(U,x):=(\Pi_h^{(k)}F_U^{(k)})(x)
 =\lambda_h^{-1}\langle v_x,\psi_h(U)\rangle_{u_h}.
 \label{eq:twosided}
\end{equation}
The inner product is a uniform sum over the auxiliary configurations $y$. Both factors are projected onto order $h$, so they can have either sign even though the original probabilities are nonnegative.

To verify Eq.~\eqref{eq:twosided}, write $(A_{k,h}g)(x)=\langle w_x,g\rangle_{u_h}$ for the average of $g$ after selecting $h$ photons uniformly from $x$. Its adjoint maps the $k$-photon output to the reduced one, $F_U^{(h)}=A_{k,h}^\dagger F_U^{(k)}$. The reduction factors in Eq.~\eqref{eq:du} imply $\Pi_hA_{k,h}^\dagger A_{k,h}\Pi_h=\lambda_h\Pi_h$. Consequently, applying $A_{k,h}$ to $\psi_h$ gives $\lambda_h$ times the order $h$ component on $k$-photon configurations, which yields Eq.~\eqref{eq:twosided}.

\subsubsection{Random grouping of mode labels}
Evaluating the inner product in Eq.~\eqref{eq:twosided} involves many choices of mode labels. To combine their contributions efficiently, place each mode independently into one of $h$ labeled groups, with equal probability~\cite{AlonColor}. Group sizes fluctuate, and empty groups are allowed. First consider configurations on $h$ distinct modes, and select those with one occupied mode in each group. Their contributions can be combined by multiplying sums over the groups and evaluating determinant generating functions.

For any fixed set of $h$ distinct modes, all $h^h$ group placements are equally likely, and $h!$ of them place one mode in each group. The set is therefore selected with probability $h!/h^h$. Weighting the selected contribution by the reciprocal of this probability recovers the original contribution on average. When the auxiliary configuration has repeated occupations, use as many groups as there are occupied modes.

For an auxiliary $h$-photon configuration $y$, let $\nu(y)=h-|\operatorname{supp}y|$, where $\operatorname{supp}y$ is the set of occupied modes, count photons beyond the first photon in each occupied mode. Repeated occupations require additional sums over their multiplicities, so we include only $\nu(y)<b$ and later choose $b$ to control their cost and omission error. This cutoff applies to auxiliary configurations used in the calculation. The final output $x$ ranges over all $k$-photon configurations, including collisions.

For each value $\nu<b$ that occurs, let $c(i)\in\{1,\ldots,h-\nu\}$ be the independently and uniformly chosen group index of mode $i$. A configuration with $\nu(y)=\nu$ has $h-\nu$ occupied modes. The probability that these lie in different groups is
\begin{equation}
 p_{h-\nu}=\frac{(h-\nu)!}{(h-\nu)^{h-\nu}}.
\end{equation}
Let $\mathcal M_h$ multiply a configuration by $1/p_{h-\nu}$ if $\nu(y)<b$ and its occupied modes lie in different groups, and set its component to zero otherwise. Let $G_h$ be the diagonal projector onto all configurations with $\nu(y)<b$. The selection correction gives $\E \mathcal M_h=G_h$. Our estimate of the contribution at order $h$ is
\begin{equation}
 \widehat f_h=\lambda_h^{-1}\langle v_x,\mathcal M_h\psi_h\rangle.
 \label{eq:random-sum}
\end{equation}
For fixed $U,x$, its mean over the random groups is the sum restricted to $\nu(y)<b$. If no configurations are omitted, $\E \widehat f_h=f_h(U,x)$. The grouping introduces variance, while the auxiliary cutoff can introduce bias. All group indices are drawn independently of $U$ and $x$.

\subsubsection{Accuracy of the random sum}
We now bound the two errors in $\widehat f_h$. The proposal $x$ is uniform and independent of $U$, and all averages are taken before the acceptance step. The error is controlled by the second moments of the two factors in Eq.~\eqref{eq:twosided}, namely the averaged outer products of $v_x$ over uniform proposals and of $\psi_h(U)$ over Haar interferometers. Both are proportional to $\Pi_h$, and they connect the error to the correlation weight $W_h(k)$,
\begin{align}
 \E_{x\sim u_k}|v_x\rangle\langle v_x|&=\lambda_h\Pi_h,
 \label{eq:effectframe}\\
 \E_U|\psi_h(U)\rangle\langle\psi_h(U)|
 &=\frac{\lambda_h W_h(k)}{Z_{M,h}}\Pi_h.
 \label{eq:sourceframe}
\end{align}
Here the outer products act on functions in $\mathcal H_h$. The first identity averages over a uniform proposal $x$, and the second over the Haar interferometer $U$. Each supplies a factor of $\lambda_h$, canceling the factor $\lambda_h^{-2}$ when we bound the squared error of Eq.~\eqref{eq:twosided}. This cancellation relies on projecting both factors onto the same correlation order, and the remaining scale is set by $W_h(k)$.

The first identity follows from $\Pi_hA_{k,h}^\dagger A_{k,h}\Pi_h=\lambda_h\Pi_h$, established above. For the second identity, let $\sigma_{N,h}^{[h]}=\mathcal P_h^{\rm op}\sigma_{N,h}$ be the order $h$ operator component of the reduced input $\sigma_{N,h}$. Haar averaging within this correlation sector, as in Appendix~\ref{app:correlations}, gives
\begin{equation}
 \E_U\ket{\psi_h}\bra{\psi_h}
 =\frac{D_h\|\sigma_{N,h}^{[h]}\|_{\HS}^2}{R_{M,h}}\Pi_h.
 \label{eq:sourceframe-expanded}
\end{equation}
The factor $D_h$ accounts for the normalization of the rescaled output probability and the uniform inner product, and $R_{M,h}$ is the operator sector dimension in Eq.~\eqref{eq:irrepdim}. Taking the trace and using the particle reduction factor gives $\E_U\|\psi_h\|_2^2=\lambda_h W_h(k)$, yielding Eq.~\eqref{eq:sourceframe}.

These identities lead to the following bounds for the random weights just constructed.

\begin{lemma}[Error from random selection]\label{lem:mask}
The following bounds apply to any diagonal projector $G_h$ onto a fixed set of auxiliary configurations and any real diagonal weights $\mathcal M_h$ that vanish outside this set and are independent of $U,x$. If $\E \mathcal M_h=G_h$ and $\E(\mathcal M_h-G_h)^2\le C_hG_h$, the estimator in Eq.~\eqref{eq:random-sum} satisfies
\begin{align}
 \E_{U,x}|f_h-\E \widehat f_h|^2
 &=\frac{W_h(k)}{Z_{M,h}}
 \|\Pi_h(I-G_h)\Pi_h\|_{\HS}^2,\label{eq:maskbias}\\
 \E_{U,x}\operatorname{Var}(\widehat f_h)
 &\le C_hW_h(k).\label{eq:maskvar}
\end{align}
The inner expectation and variance are over the random weights.
\end{lemma}
\begin{proof}
For any fixed matrix $Y$, Eqs.~\eqref{eq:effectframe} and \eqref{eq:sourceframe} give
\[
 \E_{U,x}\bigl|\lambda_h^{-1}\langle v_x,Y\psi_h\rangle\bigr|^2
 =\frac{W_h(k)}{Z_{M,h}}\|\Pi_hY\Pi_h\|_{\HS}^2.
\]
Setting $Y=I-G_h$ proves the bias identity. Setting $Y=\mathcal M_h-G_h$ and averaging over $\mathcal M_h$ gives the variance. The bound follows from $\|\Pi_hY\Pi_h\|_{\HS}^2\le\Tr(\Pi_hY^2\Pi_h)$ for Hermitian $Y$, the assumed second moment of $\mathcal M_h-G_h$, and $\Tr\Pi_h=Z_{M,h}$.
\end{proof}
For Fock state boson sampling we will also need a bound on $\E|\widehat f_h|^2$. The same proof, with $Y=\mathcal M_h$, gives $\E_{U,x,\mathcal M_h}|\widehat f_h|^2\le C'_hW_h(k)$ whenever $\E \mathcal M_h^2\le C'_hG_h$.

For the weights from random grouping, each eligible configuration has weight $1/p_{h-\nu}$ with probability $p_{h-\nu}$ and zero otherwise. Its variance is $p_{h-\nu}^{-1}-1\le \mathrm e^h$, so
\begin{equation}
 \E(\mathcal M_h-G_h)^2\le \mathrm e^hG_h.\label{eq:colormoment}
\end{equation}
Thus Lemma~\ref{lem:mask} bounds the variance averaged over $U,x$ by $\mathrm e^hW_h(k)$. The bound also holds when selection events for distinct supports are correlated.

To bound the bias, use $\|\Pi_h(I-G_h)\Pi_h\|_{\HS}^2\le\operatorname{rank}(I-G_h)$ in Eq.~\eqref{eq:maskbias}. It remains to count the omitted auxiliary configurations. Applying the insertion rule from the direct reconstruction to the empty configuration $h$ times generates a uniform $h$-photon configuration. At step $j+1$, the probability of reusing an occupied mode is at most $2j/M$. A union bound over $b$ reuse times yields
\begin{equation}
 \frac{\operatorname{rank}(I-G_h)}{Z_{M,h}}
 \le\frac{M+h-1}{M-1}
 \min\left\{1,\frac{[h(h-1)/M]^b}{b!}\right\}.
 \label{eq:repeatbias}
\end{equation}
Together, Eqs.~\eqref{eq:colormoment} and \eqref{eq:repeatbias} control the variance from random grouping and the bias from the auxiliary cutoff.

\subsection{Computing the random sum}\label{sec:contractions}
We now give an explicit procedure for evaluating $\widehat f_h$ for a supplied interferometer $U$, proposed output $x$, and fixed random groups. We first combine the projected weights into a sum of photon count moments. This sum is then evaluated with a generating function for each input family. Throughout this calculation, $V$ denotes the columns of $U$ corresponding to the nonvacuum input modes.

\subsubsection{Combining the photon count moments}\label{sec:moments}
Consider first an auxiliary configuration with one photon in each mode of a set $S$, where $|S|=h$. Its contribution is the product of the weight $v_x(S)$ of the proposed configuration and the projected output weight $\psi_h(S)$. To express both factors in terms of moments, let $\kappa_0,\ldots,\kappa_h$ solve
\begin{equation}
 \sum_{r=j}^h \kappa_r\lambda_{r,j}^{(h)}=\mathbf1_{\{j=h\}},
\end{equation}
so that $\Pi_h=\sum_r\kappa_r\mathcal T_{h,r}$ removes all orders below $h$. Applying this projector to the two factors in Eq.~\eqref{eq:framesetup} gives
\begin{equation}
 v_x(S)=\sum_{\ell=0}^{h}\alpha_\ell e_\ell(x_S),\qquad\psi_h(S)=\sum_{r=0}^{h}\beta_r\sum_{\substack{S'\subseteq S\\|S'|=r}}\left\langle\prod_{i\in S'}n_i\right\rangle_N,
\end{equation}
where
\begin{equation}
 \alpha_\ell=\frac{\kappa_\ell D_\ell}{\binom h\ell\binom k\ell},
 \qquad
 \beta_r=\frac{\kappa_rD_r}{\binom hr\binom Nr}.
 \label{eq:alphabeta}
\end{equation}
Here $e_\ell(x_S)$ sums products of $\ell$ distinct occupations in $S$, $n_i$ is the photon count in output mode $i$ before loss, and $\langle\cdot\rangle_N$ is its expectation at fixed total photon number $N$.

The sum of $v_x(S)\psi_h(S)$ over all sets $S$ with one mode from each group can now be assembled from one polynomial,
\begin{equation}
 \left\langle
 \prod_{c=1}^{h}
 \left[\sum_{i:\,c(i)=c}(1+z x_i)(1+w n_i)\right]
 \right\rangle_N.
 \label{eq:colorpoly}
\end{equation}
Each factor contributes one mode from a group. The powers of $z$ and $w$ count the occupations selected from $x$ and from the optical output, respectively. Multiply the coefficient of $z^\ell w^r$ by $\alpha_\ell\beta_r$ and sum over $\ell,r$. Dividing the result by $\lambda_hD_hp_h$ gives the contribution of these configurations with distinct modes to $\widehat f_h$.

To compute the moments in Eq.~\eqref{eq:colorpoly}, use the output generating function $\Phi(\Xi)=\langle\prod_i \xi_i^{n_i}\rangle_N$, with one variable $\xi_i$ for each output mode, collected in $\Xi=\operatorname{diag}(\xi_1,\ldots,\xi_M)$. Introduce one commuting bookkeeping variable $\omega_c$ for each group, with $\omega_c^2=0$, and substitute $\xi_i=1+\omega_{c(i)}(1+zx_i)$. A term that uses a group twice vanishes, so only $2^h$ products of the group variables remain. For each subset $\mathcal R$ of group indices, the coefficient of $\omega_{\mathcal R}=\prod_{c\in\mathcal R}\omega_c$ sums the moments involving one mode from each group in $\mathcal R$, including their factors $(1+zx_i)$. Multiply this coefficient by $\prod_{c\notin\mathcal R}\sum_{i:c(i)=c}(1+zx_i)$ to include the other groups. Weight its $z^\ell$ coefficient by $\alpha_\ell\beta_{|\mathcal R|}$ and sum over $\ell,\mathcal R$. The determinant formulas below provide $\Phi(\Xi)$ for each input family.

For configurations with repeated occupations, the corresponding projection formulas are
\begin{equation}
 v_x(y)=\sum_{\ell=0}^h\alpha_\ell\sum_{\substack{\bm\mu\le y\\|\bm\mu|=\ell}}\binom y{\bm\mu}\binom x{\bm\mu},\qquad\psi_h(y)=\sum_{r=0}^h\beta_r\sum_{\substack{\bm\mu'\le y\\|\bm\mu'|=r}}\binom y{\bm\mu'}\,\E_N\binom{\widehat n}{\bm\mu'},
 \label{eq:allmultiplicityprojection}
\end{equation}
where $\widehat n=(n_1,\ldots,n_M)$ and multiindex binomials denote products over modes. These binomial moments are the coefficients of $\Phi(\Xi)$ expanded about $\xi_i=1$.

For each $\nu<b$, explicitly sum over the mode labels with occupations greater than one and their multiplicities $m_i=\nu_i+1\ge2$, with $\sum_i \nu_i=\nu$. Use $h-\nu$ groups and omit a choice if two of these modes belong to the same group. Each repeated mode is assigned its own variable, with powers above $m_i$ discarded. Multiply its moment coefficient of degree $\mu'_i$ by $\binom{m_i}{\mu'_i}$, and use $\sum_{\mu=0}^{m_i}\binom{m_i}{\mu}\binom{x_i}{\mu}z^\mu$ for the factor of the proposed configuration. The modes occupied once are summed over the unused groups as above. The variables for repeated modes require at most $\prod_i(\nu_i+2)\le3^\nu$ coefficients. Section~\ref{sec:repeat-count} bounds the number of explicit choices of mode labels and multiplicities. Each grouped contribution is divided by $\lambda_hD_hp_{h-\nu}$, as prescribed by Eq.~\eqref{eq:random-sum}.

\subsubsection{Fock state boson sampling}
For Fock state boson sampling, the reduced input $\gamma_h$ is a uniform mixture over $h$-element subsets of the $n$ occupied inputs. To compute its generating function, divide the occupied inputs into $h$ random groups independently of the output grouping. Let $d(a)$ be the group index of occupied input $a$, chosen independently and uniformly from $1,\ldots,h$, and form
\begin{equation}
 \gamma_{h,d}=\frac1{\binom nh p_h}
 \sum_{\substack{S\subseteq[n],\ |S|=h\\\text{one input per group}}}
 \ket{1_S}\bra{1_S}.
 \label{eq:inputmask}
\end{equation}
Each subset is selected with probability $p_h$, so averaging over the input groups restores $\gamma_h$. Orthogonality of the Fock states also gives
\begin{equation}
 \E_d\gamma_{h,d}=\gamma_h,\qquad
 \E_d\Tr\gamma_{h,d}^2=\frac1{p_h\binom nh}.
 \label{eq:inputmaskmoments}
\end{equation}
The number of selected subsets fluctuates with the grouping, so $\Tr\gamma_{h,d}$ equals one only on average. We therefore keep $\gamma_{h,d}$ unnormalized, since dividing each realization by its trace would bias the average and break the first identity in Eq.~\eqref{eq:inputmaskmoments}.

For single-photon inputs, $V$ has $n$ columns. Introduce input group variables $t_j$ with $t_j^2=0$, independent of the output variables $\omega_c$. With $\Theta_d=\operatorname{diag}(t_{d(1)},\ldots,t_{d(n)})$, the output generating function of $\gamma_{h,d}$, in the output variables $\Xi$ of Sec.~\ref{sec:moments}, is
\begin{equation}
 \Phi_{h,d}(\Xi)=
 \frac{[t_1\cdots t_h]\det(I-\Theta_dV^\dagger\Xi V)^{-1}}
 {\binom nh p_h}.
 \label{eq:macmahon}
\end{equation}
The coefficient notation $[t_1\cdots t_h]$ selects one variable from each input group. By MacMahon's master theorem~\cite{Tuite}, this coefficient sums the permanents of the $S\times S$ submatrices of $V^\dagger\Xi V$ over subsets $S$ containing one input from each group, exactly the weighted output sums of their Fock states. Thus a single evaluation of Eq.~\eqref{eq:macmahon} accounts for all selected input subsets together.

Substituting the output weights from the preceding subsection supplies the moments needed for $\widehat f_h$. Because $\gamma_{h,d}$ already contains $h$ photons, the source coefficients $\beta_r$ in this step use $N=h$. The input and output group variables, the variables for repeated modes, and the powers of $z$ require $2^{O(h+\nu)}\operatorname{poly}(h)$ coefficients. The inverse determinant is evaluated by the finite expansion $\exp[\sum_{j=1}^h\Tr((\Theta_dV^\dagger\Xi V)^j)/j]$, with terms that use a group twice discarded. The cost for each fixed choice of repeated modes and their occupations is therefore $\operatorname{poly}(M,n,h)2^{O(h+\nu)}$.

It remains to control the extra randomness from grouping the inputs. The second moment argument following Lemma~\ref{lem:mask} applies to any positive operator $X$ on the $h$-photon space, with arbitrary trace. Substitute the rescaled output weights $D_h\langle y|\mathcal U_hX\mathcal U_h^\dagger|y\rangle$ into that argument. Equations~\eqref{eq:sourceframe-expanded} and \eqref{eq:twosided} give the mean squared reconstructed contribution
\begin{equation}
 \mathcal W_h(X)=
 \frac{D_hZ_{M,h}}{R_{M,h}\lambda_h}
 \|\mathcal P_h^{\rm op}X\|_{\HS}^2.
 \label{eq:unnormalizedenergy}
\end{equation}
The prefactor obeys
\begin{equation}
 \frac{D_hZ_{M,h}}{R_{M,h}\lambda_h}
 =\binom kh\frac{M+h-1}{M+2h-1}
 \frac{(M+2h-1)!(M+k-1)!}{(M+h-1)!(M+k+h-1)!}
 \le\binom kh,
 \label{eq:fockframeprefactor}
\end{equation}
and orthogonal projection gives $\mathcal W_h(X)\le\binom kh\Tr X^2$. The independent output grouping contributes a factor at most $\mathrm e^h$ to $\E|\widehat f_h|^2$. Setting $X=\gamma_{h,d}$ and averaging with Eq.~\eqref{eq:inputmaskmoments}, using $p_h^{-1}\le \mathrm e^h$, yields
\begin{equation}
 \E_{U,x,c,d}|\widehat f_h|^2
 \le \mathrm e^{2h}\frac{\binom kh}{\binom nh}.
 \label{eq:fockcolorvariance}
\end{equation}
The same bound holds for empty input groups and groups of unequal size. Since the input average is exact, the only bias remains the omission of auxiliary output configurations with $\nu(y)\ge b$. For each correlation order and each proposed output, the input and output groups are drawn anew, independently of each other and of all other draws. This independence makes the variances of different orders add, as used in Sec.~\ref{sec:combined-error}.

\subsubsection{Gaussian boson sampling}\label{sec:gaussian-evaluation}
For Gaussian boson sampling, the output generating function is available directly as a determinant~\cite{Fitzke2023}. A pure Gaussian output state with no displacement has the form~\cite{Hamilton}
\begin{equation*}
 \ket{\psi_B}=\mathcal N_B
 \exp\!\left(\frac12\sum_{i,j=1}^{M}B_{ij}a_i^\dagger a_j^\dagger\right)\ket0,
\end{equation*}
where $a_i^\dagger$ creates a photon in output mode $i$ and $\mathcal N_B$ normalizes the state. The matrix $B$ of pair creation coefficients is complex symmetric, $B^T=B$. For $s$ equally squeezed inputs, $V$ has $s$ columns and $B=(\tanh r)H$, with $H=VV^T$. In particular, $H=UU^T$ when all inputs are squeezed.

To condition this state on $N$ photons, extract the coefficient of $q^{N/2}$ from its counting determinant and divide by the same coefficient at $\Xi=I$. Writing $[q^m]f$ for the coefficient of $q^m$ in $f$, the resulting output generating function is
\begin{equation}
 \Phi(\Xi)= \frac{[q^{N/2}]\det(I-qB^\dagger\Xi B\Xi)^{-1/2}} {\mathcal Z_N},\qquad \mathcal Z_N=[q^{N/2}]\det(I-qB^\dagger B)^{-1/2}.\label{eq:gaussiancolor}
\end{equation}
This is the function required in the common construction above. Substituting the group variables and combining the coefficients as in Eq.~\eqref{eq:colorpoly} evaluates the grouped moments. The common squeezing scale cancels between the numerator and denominator at fixed $N$, so the calculation may use $H$ in place of $B$.

The determinant can be computed as a finite series. Write $\det(I-qB^\dagger\Xi B\Xi)^{-1/2}=\sum_{j\ge0}g_jq^j$. Its coefficients satisfy
\begin{equation}
 g_0=1,\qquad
 g_j=\frac1{2j}\sum_{\ell=1}^{j}\Tr[(B^\dagger\Xi B\Xi)^\ell]g_{j-\ell}.
\end{equation}
Only $j\le N/2$ is needed. Before including $q$, the group variables and $z$ require $O(2^h(h+1))$ coefficients. Multiplying them costs $O(3^h\operatorname{poly}(h))$ operations because only disjoint subsets of groups contribute. The cost is therefore $\operatorname{poly}(M,N,h)2^{O(h)}$ for distinct auxiliary modes. Including variables for repeated modes gives $\operatorname{poly}(M,N,h)2^{O(h+\nu)}$ for each fixed choice of repeated modes and their occupations, as in the single-photon calculation.

\subsection{Evaluation at the requested accuracy}
We now combine the estimates at different correlation orders and choose the auxiliary cutoff $b$. The goal is to keep the bias from omitted auxiliary configurations at most $\epsilon/16$, its share of the total error budget in Appendix~\ref{app:sampling}, with evaluation cost polynomial in the system size and $1/\epsilon$.

\subsubsection{Combining the correlation orders}\label{sec:combined-error}
Using independent random groups at different orders, form $\widehat F=1+\sum_{h=1}^{\min\{L,k\}}\widehat f_h$ as the estimate of $F_U^{\le L}(x)$. We control both the bias caused by omitting auxiliary configurations with $\nu(y)\ge b$ and the variance from random grouping. The truncation of correlation orders above $L$ is a separate error, already bounded in Appendix~\ref{app:correlations}.

The bounds below average over the photon sectors in which the estimator is used, each weighted by its original probability. For $h\ge0$, let $H_h$ be the resulting average of $W_h(k)$, with $W_0(k)=1$ and $W_h(k)=0$ for $h>k$, so that $H_0$ is the total probability of these sectors. We fix a constant $C_2\ge\max\{1,\sum_{h\ge0}H_h\}$, which bounds the averaged second moment of $F_U$, and write $S_L=\sum_{h>L}H_h$ for the correlation tail. Averages below are taken over Haar random $U$, the photon sectors and the uniform proposal $x$ within each sector. By orthogonality, $S_L$ is then the mean squared error from omitting orders above $L$. Finally, let $B_{\rm aux}$ be the root mean squared bias of $\widehat F$ relative to $F_U^{\le L}$, and let $V_L$ be its variance over the random groups, written $V_L^{\mathrm F}$ and $V_L^{\mathrm G}$ for the two input families.

At each order $h$, the bias $\E\widehat f_h-f_h$ is a function of $x$ of correlation order $h$, because $v_x$ is the order $h$ part of $w_x$. Biases at distinct orders are therefore orthogonal under the uniform average over $x$. Their squared norms add, and Eq.~\eqref{eq:repeatbias} gives
\begin{equation}
 B_{\rm aux}^2
 \le\sum_{h=1}^{L}H_h
 \frac{M+h-1}{M-1}
 \min\left\{1,\frac{[h(h-1)/M]^b}{b!}\right\}.
 \label{eq:totalbias}
\end{equation}
Independence of the groups at different orders makes the conditional variances add as well. For Gaussian boson sampling, the bound $\mathrm e^hW_h(k)$ at order $h$ yields
\begin{equation}
 V_L^{\mathrm G}\le \mathrm e^LC_2.
 \label{eq:totalvariance}
\end{equation}
For Fock state boson sampling, grouping the inputs introduces the additional factor in Eq.~\eqref{eq:fockcolorvariance}. Since $k\sim\operatorname{Bin}(n,\eta)$ and $\E_k\binom kh=\binom nh\eta^h$, summing over orders gives
\begin{equation}
 V_L^{\mathrm F}\le\sum_{h=1}^{L}\mathrm e^{2h}\eta^h\le\frac{\mathrm e^{2L}}{1-\eta}.
 \label{eq:focktotalvariance}
\end{equation}

\subsubsection{Counting the modes with repeated occupations}\label{sec:repeat-count}
The explicit sum over repeated occupations in Sec.~\ref{sec:contractions} runs over choices of which modes have occupations greater than one and what those occupations are. At order $h$, one such choice consists of distinct labels $i_1<\cdots<i_j$ and occupations $\nu_1+1,\ldots,\nu_j+1\ge2$, with $\sum_a \nu_a=\nu$. This choice uses $j+\nu$ photons. The remaining $h-j-\nu$ photons occupy distinct modes and are handled by the grouped sum.

For $\nu\ge1$, there are $\binom Mj$ ways to choose the $j$ labels and $\binom{\nu-1}{j-1}$ ways to distribute $\nu$ extra photons among them, with at least one extra photon per mode. Ignoring the restriction $j+\nu\le h$ gives the upper bound
\begin{equation}
 \sum_{j=1}^{\min(M,\nu)}\binom Mj\binom{\nu-1}{j-1}
 =\binom{M+\nu-1}{\nu}.
 \label{eq:assigncount}
\end{equation}
There is one empty choice at $\nu=0$. Let $K_b$ count the choices with $\nu<b$ at the largest order $L$. This also bounds their number at every $h\le L$. Summing Eq.~\eqref{eq:assigncount}, including $\nu=0$, gives
\begin{equation}
 K_b\le\binom{M+b-1}{b-1}.
 \label{eq:assignsum}
\end{equation}
Since $\nu\le L-1$, the sum can always be stopped at $\min\{b-1,L-1\}$.

For each such choice, the calculation of Sec.~\ref{sec:contractions} uses $h-\nu$ group variables and a variable of degree at most $\nu_a+1$ for each repeated mode. Including the powers of $z$, the number of coefficients is at most
\begin{equation}
 2^{h-\nu}\prod_a(\nu_a+2)\operatorname{poly}(h)
 \le2^h3^\nu\operatorname{poly}(h).
\end{equation}
For Fock state boson sampling, the input groups add at most $2^h$ products of variables. The cost per choice is therefore polynomial in $M,N,h$ and exponential in $h+\nu$. To bound the full cost, it remains to control $K_b$ while making the omission bias small.

\subsubsection{Choosing the auxiliary cutoff}\label{app:colorcost}
If $L=0$, the estimator is $\widehat F=1$, with zero auxiliary bias and variance and no auxiliary enumeration. For $L\ge1$, increasing $b$ reduces the bias in Eq.~\eqref{eq:totalbias} and raises the number $K_b$ of explicit choices. We balance these two effects through the scale
\begin{equation}
 L_{\rm eff}=L+1+\left\lceil\log_2\frac{40(1+C_2)}{\epsilon}\right\rceil.
 \label{eq:budgetS}
\end{equation}
We will keep $K_b=2^{O(L_{\rm eff})}$, which is polynomial in $1/\epsilon$ when $C_2=O(1)$ and $L=O(1+\log(1/\epsilon))$.

If $M\le64L_{\rm eff}$, include every repeated occupation by choosing $b=L$. Then $B_{\rm aux}=0$, and Eq.~\eqref{eq:assignsum} gives
\begin{equation}
 K_b\le\binom{M+L-1}{L-1}\le2^{M+L}\le2^{65L_{\rm eff}}.
\end{equation}
If $M>64L_{\rm eff}$, repeated occupations are rare under the uniform distribution of auxiliary configurations. Set $d_*=\lfloor\log_2(M/L_{\rm eff})\rfloor\ge6$ and choose $b=\lceil8L_{\rm eff}/d_*\rceil$. If $b\ge L$, all configurations are again included. Otherwise, applying the counting bound behind Eq.~\eqref{eq:repeatbias} and $b!\ge(b/\mathrm e)^b$ gives, for each order $h\le L$,
\begin{equation}
 \Pr_{u_h}(\nu\ge b)
 \le\left(\frac{\mathrm e L^2}{Mb}\right)^b
 \le\left(\frac{\mathrm e d_*}{8\,2^{d_*}}\right)^b
 \le2^{-4L_{\rm eff}}.
\end{equation}
The middle inequality uses $L\le L_{\rm eff}$, $M\ge2^{d_*}L_{\rm eff}$, and $b\ge8L_{\rm eff}/d_*$. For the last inequality, $d_*\log2-\log d_*+\log8-1\ge(d_*/2)\log2$ for $d_*\ge6$. The remaining prefactor $(M+h-1)/(M-1)$ in Eq.~\eqref{eq:totalbias} is less than two, so
\begin{equation}
 B_{\rm aux}^2\le2C_2\,2^{-4L_{\rm eff}}\le(\epsilon/16)^2.
\end{equation}

To bound the cost in the same regime, let $\nu_{\max}=\min\{b-1,L-1\}<8L_{\rm eff}/d_*$. If $\nu_{\max}=0$, there is only the empty choice. Otherwise, the factorial bound used above, $M<2^{d_*+1}L_{\rm eff}$, and monotonicity of $x\log[\mathrm e(M+x)/x]$ give
\begin{equation}
 \log_2 K_b\le\log_2\binom{M+\nu_{\max}}{\nu_{\max}}\le
 \nu_{\max}\log_2\frac{\mathrm e(M+\nu_{\max})}{\nu_{\max}}\le12L_{\rm eff}.
\end{equation}
Hence, for $M>64L_{\rm eff}$,
\begin{equation}
 B_{\rm aux}\le\epsilon/16,\qquad K_b\le2^{12L_{\rm eff}}.
 \label{eq:adaptivecost}
\end{equation}
Both regimes therefore achieve the required bias with $K_b=2^{O(L_{\rm eff})}$. Multiplying by the cost per choice and summing over $h\le L$ gives, for both input families,
\begin{equation}
 T_{\rm eval}\le\operatorname{poly}(M,N,L)\,\mathrm e^{O(L_{\rm eff})},\qquad L_{\rm eff}=O\!\left(L+\log(1+C_2)+\log(1/\epsilon)\right).\label{eq:fastcost}
\end{equation}
The separate treatment of $M\le64L_{\rm eff}$ makes this bound uniform in the mode count and the requested accuracy.

\section{Construction of the sampler}\label{app:sampling}
Appendix~\ref{app:correlations} bounds the truncation error, and Appendix~\ref{app:evaluation} constructs estimators of $F_U^{\le L}(x)$ with controlled error and evaluation cost. We now turn these estimates into the sampling algorithms of Theorems~\ref{thm:fock} and \ref{thm:gaussian}. We average and clip them to obtain acceptance probabilities and bound the number of rejection trials. Bounds on photon number tails and on finite precision complete the error and running time analysis.

\subsection{Rejection sampling}\label{sec:sampler}
For a fixed $U$ and photon sector, write $F=F_U$ and let $u$ be the uniform occupation law, so that the target law is $p_F(x)=u(x)F(x)$. The truncated expansion and its random estimate can have negative values. To obtain a nonnegative weight with a chosen upper bound $A>0$, we use clipping,
\[
 \clip_{[0,A]}(z)=\min\{A,\max\{0,z\}\}.
\]
This replaces values below zero by zero and values above $A$ by $A$, leaving the others unchanged. Dividing the resulting weight by $A$ gives a valid acceptance probability. We use this probability in the standard rejection method~\cite{Devroye1986}, with a finite trial cutoff.

To bound the output error, we first fix the acceptance weight.

\begin{lemma}[Finite rejection]\label{lem:rejection}
Let $F\ge0$ with $\E_uF=1$, let $0\le a\le A$ be a fixed acceptance weight, and put $\Delta=\E_u|a-F|$. Consider the sampler that proposes $x\sim u$, accepts with probability $a(x)/A$, and after $J$ rejected trials outputs a fixed configuration with the same total. Its output distribution $P_J$ obeys
\begin{equation}
 \TV(P_J,p_F)\le2\Delta+\frac{A}{\mathrm eJ}.
 \label{eq:retry}
\end{equation}
The bound remains valid when the acceptance probability vanishes.
\end{lemma}
\begin{proof}
Let $m_a=\E_ua$. If $m_a>0$, the accepted law is $u(x)a(x)/m_a$ and its distance from $p_F$ is at most $\Delta$, since $|m_a-1|\le\Delta$. Failure contributes $(1-m_a/A)^J$. This is at most $\Delta+m_a(1-m_a/A)^J$, and the latter product is bounded by $A/(\mathrm eJ)$. The same bound holds at $m_a=0$, when $\Delta=1$.
\end{proof}

For the estimator from Appendix~\ref{app:evaluation}, let $\widehat F_T(x)$ be the average of $T$ fresh independent copies at the proposed $x$. The sampler accepts with probability $\clip_{[0,A]}(\widehat F_T(x))/A$. For the error analysis, averaging over the internal estimator randomness gives the weight $a(x)=\E[\clip_{[0,A]}(\widehat F_T(x))]$ in Lemma~\ref{lem:rejection}.

Clipping does not increase distances, $|\clip_{[0,A]}(z)-\clip_{[0,A]}(z')|\le|z-z'|$. Since $F\ge0$, so that $\clip_{[0,A]}(F)=\min\{F,A\}$, this implies
\[
 |\clip_{[0,A]}(z)-F|\le |z-F|+(F-A)_+,
 \qquad t_+:=\max\{t,0\}.
\]
The first term accounts for the approximation error, and the second for the tail removed by the upper bound $A$. To combine photon sectors, first use their original probabilities to form a distribution with the approximate sampler in each sector where it is used and the target law in all other sectors. Its error is bounded by the weighted sector errors. Replacing exceptional sectors by fixed outputs adds at most their total probability, and restricting the photon count distribution adds at most the omitted probability. These contributions form $\epsilon_{\rm count}$.

For the full output law $Q_U$, it remains to bound $\Delta=\E_u|a-F|$ in Lemma~\ref{lem:rejection}, averaged as in Sec.~\ref{sec:combined-error}. By the clipping bound and the triangle inequality, $\Delta$ splits into four parts: the truncation error $|F-F_U^{\le L}|$, the bias of $\widehat F$, the fluctuation of $\widehat F_T$ around its mean, and the tail $(F-A)_+$. By the Cauchy--Schwarz inequality, their averages are at most $\sqrt{S_L}$, $B_{\rm aux}$, and $\sqrt{V_L/T}$, and $(F-A)_+\le F^2/A$ bounds the last by $C_2/A$. Inserting these into Lemma~\ref{lem:rejection} and adding the sector errors gives
\begin{equation}
 \E_U\TV(Q_U,P_U)\le{} 2\sqrt{S_L}+2B_{\rm aux}+2\sqrt{V_L/T} +\frac{2C_2}{A}+\frac{A}{\mathrm eJ} +\epsilon_{\rm count}+\epsilon_{\rm num}.\label{eq:masterTV}
\end{equation}
Numerical errors and errors from finite random bits are collected in $\epsilon_{\rm num}$ and bounded in Sec.~\ref{app:precision}.

Choose $\sqrt{S_L}\le\epsilon/16$, use the auxiliary cutoff from Sec.~\ref{app:colorcost}, and take
\begin{equation}
 T\ge256V_L/\epsilon^2,\quad
 A\ge16C_2/\epsilon,\quad
 J\ge16A/\epsilon.
 \label{eq:algorithmparameters}
\end{equation}
In the regimes of Theorems~\ref{thm:fock} and \ref{thm:gaussian}, $C_2=O(1)$ and $S_L$ decays exponentially in $L$. For single-photon inputs, Eq.~\eqref{eq:focktail} gives $\sqrt{S_L}\le\epsilon/16$ whenever
\begin{equation}
 L+1\ge\frac{\log[256/((1-\eta)\epsilon^2)]}{-\log\eta},
 \label{eq:lowlosscut}
\end{equation}
and Eq.~\eqref{eq:alltail} gives the same condition for Gaussian inputs with $L=2\ell$, $\ell$ in place of $L$, and $\chi_*^2$ in place of $\eta$. In both cases $L=O(1+\log(1/\epsilon))$, with constants that depend on $\eta$ and, for the Gaussian theorem, on $r$. By Eqs.~\eqref{eq:fastcost}, \eqref{eq:totalvariance}, and \eqref{eq:focktotalvariance}, every factor is then polynomial in system size and $1/\epsilon$.

\subsection{Prefix sampling for small photon sectors}\label{app:prefix}
For sectors containing few photons, we evaluate prefix probabilities, the probabilities of the photon counts in the first few output modes, and sample sequentially, following the construction used in exact Gaussian boson sampling algorithms~\cite{QuesadaExact}. The coefficient calculation below additionally fixes the input photon sector.

Use the matrix $B$ of pair creation coefficients introduced in Sec.~\ref{sec:gaussian-evaluation}. Separate the output modes into the set $S$ whose photon counts have been specified and the remaining set $R$, and write
\begin{equation}
 B=\begin{pmatrix}B_{SS}&B_{SR}\\B_{SR}^T&B_{RR}\end{pmatrix}.
\end{equation}
The Bargmann representation~\cite{Bargmann1961} writes a photon number state $\ket{m}$ as $t^m/\sqrt{m!}$, with one variable per mode. A density operator then becomes a generating function of its photon number matrix elements, called its Bargmann kernel, with separate variables for the ket and bra indices~\cite{JohnParthasarathy2021}. Let $\boldsymbol u$ and $\boldsymbol v$ denote these two sets of variables on $S$. Summing over the photon counts in $R$, with a factor $z$ for each photon, gives the unnormalized kernel
\[
 \mathcal K_z(\boldsymbol u,\boldsymbol v)=\mathcal Z_R(z)\exp\!\left[
 \tfrac12\boldsymbol u^T\mathcal E(z)\boldsymbol u+
 \boldsymbol u^T\mathcal F(z)\boldsymbol v+
 \tfrac12\boldsymbol v^T\overline{\mathcal E(z)}\boldsymbol v\right],
\]
where
\begin{align}
 \mathcal Z_R(z)&=\det(I-z^2B_{RR}^\dagger B_{RR})^{-1/2},\\
 \mathcal E(z)&=B_{SS}+z^2B_{SR}(I-z^2B_{RR}^\dagger B_{RR})^{-1}B_{RR}^\dagger B_{SR}^T,\\
 \mathcal F(z)&=zB_{SR}(I-z^2B_{RR}^\dagger B_{RR})^{-1}B_{SR}^\dagger.
\end{align}
For $\mathcal E(z)=\sum_{j\ge0}\mathcal E_jz^j$, the notation $\overline{\mathcal E(z)}$ means $\sum_{j\ge0}\overline{\mathcal E_j}z^j$. For a configuration $x$ on $S$ with $|x|\le N$, its probability conditioned on $N$ total photons is
\begin{equation}
 \Pr(x\mid N)=
 \frac{[z^{N-|x|}]\left\{\mathcal Z_R(z)\,\frac{1}{x!}
 \operatorname{haf}\mathcal H_x(z)\right\}}{\mathcal Z_N},
 \label{eq:prefixprob}
\end{equation}
where
\begin{equation}
 \mathcal H_x(z)=
 \begin{pmatrix}\mathcal E(z)&\mathcal F(z)\\\mathcal F(z)^T&\overline{\mathcal E(z)}\end{pmatrix}_{(x,x)}.
\end{equation}
The subscript $(x,x)$ repeats the row and column for mode $i$ exactly $x_i$ times in each of the two index sets. To obtain Eq.~\eqref{eq:prefixprob}, differentiate $\mathcal K_z$ exactly $x_i$ times in each of $u_i$ and $v_i$, and evaluate at $\boldsymbol u=\boldsymbol v=0$. These $2|x|$ derivatives of a quadratic exponential sum products over all pairings, giving the hafnian of $\mathcal H_x(z)$. Division by $x!$ accounts for the normalization of the photon number states. Extracting $[z^{N-|x|}]$ selects $N-|x|$ photons in $R$, and division by $\mathcal Z_N$ normalizes the $N$ photon sector.

The hafnian has size $2|x|\le2N$, and its entries are power series truncated at degree $N$. The hafnian algorithm of Ref.~\cite{BjorklundHafnian} applies to such entries. We evaluate it by dynamic programming over subsets and form, from successive prefix probabilities, the finite conditional law of the next count $x_i$, which ranges from zero to the number of photons not yet placed. The prefix calculations cost $\operatorname{poly}(M,N)4^N$ arithmetic operations per sample. At finite binary matrix inputs, the weights are evaluated by exact rational arithmetic and prefixes of zero probability are omitted. The conditional laws are sampled with finite random bits, with their error allocated to $\epsilon_{\rm num}$. After the configuration of all $N$ photons is drawn, uniform hypergeometric deletion of $N-k$ photons applies the loss.

\subsection{Photon number tails at finite dimension}\label{app:counts}
For Gaussian boson sampling, the input photon number $N$ takes the values $0,2,4,\ldots$ with the probabilities in Eq.~\eqref{eq:countlaw}. We choose a threshold $N_0$ and use prefix sampling for $N\le N_0$, applying the correlation approximation to larger sectors where the bounds of Appendix~\ref{app:correlations} hold. The bounds below determine $N_0$ and control the probability outside the count region where the approximation is used, uniformly in the number of squeezed inputs.

For $s$ equally squeezed modes, write $q_r=\tanh^2r$ and $c_0=\sinh^2r$. The following count estimates hold for every $r>0$ and every $\eta$, uniformly in $1\le s\le M$. Choose constants
\begin{equation}
 c_0<c_1<c,
 \qquad \eta<\vartheta<1.
\end{equation}
Choose $z>1$ close enough to one that $q_rz^2<1$ and
\begin{equation}
 R_r(z):=\frac{1-q_r}{1-q_rz^2}\le z^{2c_1}.
\end{equation}
Such a $z$ exists because the logarithmic derivative of $R_r$ at one is $2c_0<2c_1$. The input law obeys
\begin{equation}
 \E z^N=R_r(z)^{s/2}\le z^{c_1s}.
\end{equation}
Let $\alpha=1-c_1/c>0$. Splitting into $s\le N_0/c$ and $s>N_0/c$ yields
\begin{align}
 \Pr(N>N_0,\ N>cs)&\le z^{-\alpha N_0},\label{eq:densetail}\\
 \Pr(N>N_0,\ k>\vartheta N)
 &\le \mathrm e^{-2(\vartheta-\eta)^2N_0}.\label{eq:binomialtail}
\end{align}
The second bound follows by applying Hoeffding's inequality~\cite{Hoeffding1963} to $k\sim\operatorname{Bin}(N,\eta)$ conditional on $N$. Choose even $N_0=O_{r,\eta}(1+\log(1/\epsilon))$ so that each bound is at most $\epsilon/64$.

To sample from a finite photon number distribution, define $N_{\max}$ as the smallest even integer at least $cs+N_0$ and omit counts above it. The moment bound above gives $\Pr(N>N_{\max})\le z^{c_1s-N_{\max}}\le z^{-N_0}$.

Generate the input count conditioned on $N\le N_{\max}$, followed by the binomial output count. For $N\le N_0$, use the prefix sampler of Sec.~\ref{app:prefix} and then apply hypergeometric loss. For $N>N_0$ outside $N\le cs$ and $k\le\vartheta N$, emit a fixed configuration. For $N>N_0$ satisfying both bounds, use the sampler of Sec.~\ref{sec:sampler}, whose error obeys Eq.~\eqref{eq:masterTV}. As in the mixture argument preceding Eq.~\eqref{eq:masterTV}, count truncation contributes at most $\Pr(N>N_{\max})$, and exceptional sectors contribute their original probability. Since $0<\alpha<1$, the count tail also obeys $z^{-N_0}\le z^{-\alpha N_0}\le\epsilon/64$. Together with Eqs.~\eqref{eq:densetail} and \eqref{eq:binomialtail}, this gives $\epsilon_{\rm count}\le3\epsilon/64$ uniformly in $s$.

With this choice of $N_0$, the prefix calculations have polynomial cost in $M$ and $1/\epsilon$. At fixed $r$, $\eta$, and $\epsilon$, the threshold is independent of $M$.

In Theorem~\ref{thm:gaussian}, $s=M$. On $N\le cM$ and $k\le\vartheta N$, Eqs.~\eqref{eq:alltail} and \eqref{eq:allsecond} give the exponentially decaying correlation tail and the bounded second moment required by the approximate sampler, for every choice $\sinh^2r<c_1<c$ and $\eta<\vartheta<1$. Together with the count bounds above and prefix sampling for $N\le N_0$, this covers every finite $M$ at fixed $r>0$ and $\eta$.

To apply the bound of Theorem~\ref{thm:gaussian} to every smaller squeezing parameter, write $r_{\max}$ for the squeezing parameter fixed there and choose $\sinh^2r_{\max}<c_1<c$. Choose $z>1$ by the preceding argument at $r=r_{\max}$. Since $R_r(z)\le R_{r_{\max}}(z)$ for $0\le r\le r_{\max}$, the same count bounds and threshold $N_0$ hold throughout this interval. The same $c$ and $\eta<\vartheta<1$ give a uniform geometric factor $\chi_*=\vartheta(1+c)/(1+\vartheta c)<1$ in Eqs.~\eqref{eq:alltail} and \eqref{eq:allsecond}. The sector evaluator is independent of $r$, and Sec.~\ref{app:precision} controls the rounding of $r$. Thus the running time is polynomial with constants depending only on $r_{\max}$ and $\eta$. At $r=0$ the output is vacuum.

\subsection{Finite precision and bit complexity}\label{app:precision}
The theorems count arithmetic operations, and we now show that the same bounds hold in bit complexity. Here the classical input is a matrix $\widetilde U$ given by finite binary approximations to the real and imaginary parts of $U$, each with absolute error at most $2^{-p}$, at a precision $p$ prescribed by the algorithm. This is the usual finite precision convention for matrix inputs~\cite[Sec.~2]{AaronsonArkhipov}, and the target distribution remains that of the ideal $U$. The fixed optical parameters are computable to absolute error $2^{-p}$ in time polynomial in $p$. In the statement uniform over $0\le r\le r_{\max}$, $r$ is instead supplied at the prescribed precision, and the target is defined by its ideal value. For single-photon inputs, set $N_{\max}=n$, and for Gaussian inputs use the photon count cutoff above. Allocate $\epsilon_{\rm num}=\epsilon/4$ to matrix rounding, scalar calculations and finite random bits in Eq.~\eqref{eq:masterTV}. We show that $p=\operatorname{poly}(M+N_{\max}+L+\log(1/\epsilon_{\rm num}))$ bits per matrix component suffice.

At finite binary matrix inputs, the coefficient calculations use exact rational arithmetic. Their degrees in the matrix entries and conjugates are polynomial in $N_{\max}+L$. The binary input denominators, determinant recurrence factors, and rational projection and grouping coefficients admit a common denominator of polynomial bit length in $M+N_{\max}+L+p$. The logarithm of the number of expanded terms is polynomial in the cutoffs, so the numerators also have polynomial bit length. Normalizing a rounded photon sector additionally introduces its rational normalization coefficient, whose numerator and denominator satisfy the same bound. Each arithmetic operation therefore has polynomial bit cost.

These expansions also control matrix rounding. A degree $d$ monomial in real and imaginary matrix components changes by at most $d2^{d-1}2^{-p}$, since ideal components have magnitude at most one and rounded components at most two. Multiplying by the sum of absolute coefficients bounds each polynomial's change. The degrees and logarithms of these sums are polynomially bounded in the cutoffs, uniformly over proposed configurations and group assignments. Choosing $p$ sufficiently large therefore makes the acceptance errors, summed over the bounded number of trials, smaller than their numerical budget. Clipping cannot increase these errors.

For equal squeezing, the common factor in $B=(\tanh r)H$ cancels within each fixed photon sector. We evaluate the coefficients using $H$. For the ideal $U$, the sector normalization is
\[
 \frac{1}{2^{N/2}(N/2)!}\prod_{\ell=0}^{N/2-1}(s+2\ell).
\]
The approximate estimator retains this known normalization. For prefix sampling, we instead normalize the state of $N$ photons generated by the rounded matrix before applying loss. Put $\delta_{\rm op}=\|\widetilde U-U\|_{\rm op}\le1/2$. Every singular value of $\widetilde U$ is at least $1/2$, so its induced map on the $N$-photon sector has minimum singular value at least $2^{-N}$. The rounded sector normalization is therefore at least $2^{-2N}$ times the ideal coefficient above. That coefficient is at least $2^{-N/2}$ for $s\ge1$, so normalization requires only $O(N)$ additional precision bits.

A telescoping sum over the $N$ tensor factors bounds the change of a normalized vector in an input sector before renormalization by $(1+\delta_{\rm op})^N-1$. After renormalization, its distance from the ideal output vector is at most $2[(1+\delta_{\rm op})^N-1]$. Photon loss and detection contract trace distance, so this also bounds the output total variation error. Since $\delta_{\rm op}\le\sqrt2\,M2^{-p}$, taking $p=O(\log(2+MN_{\max}/\epsilon_{\rm num}))$ sufficiently large controls matrix rounding in the prefix branch. These bounds remain uniform as $r$ approaches zero.

Rounding the squeezing parameter is controlled in the same way. The squared overlap of the product squeezed vacuum inputs at $r$ and $r'$ is $\cosh(r-r')^{-s}$. Their trace distance is at most $\sqrt{s/2}\,|r-r'|$, using $\log\cosh t\le t^2/2$ and $1-\mathrm e^{-x}\le x$ for $x\ge0$. The optical channel cannot increase this distance. Hence $O(\log(M/\epsilon_{\rm num}))$ bits of absolute precision suffice for the supplied squeezing throughout any fixed bounded interval, with the resulting error included in $\epsilon_{\rm num}$.

The photon count weights are obtained from Eq.~\eqref{eq:countlaw} by dropping the common factor $\sech^s r$ and normalizing over $N\le N_{\max}$. The vacuum weight is one. Evaluating these finite weights and the binomial loss probabilities to sufficient precision has polynomial cost under the input convention above.

Finally, sampling a finite law on a set $\mathcal X$ of outcomes using a $p_{\rm rand}$ bit uniform grid incurs total variation error at most $|\mathcal X|2^{-p_{\rm rand}}$. This applies to group labels, integer ranks of configurations, conditional photon counts, and binary acceptance decisions. The number of choices and the logarithms of their outcome counts are polynomially bounded. Choosing their errors to sum to the remaining $\epsilon_{\rm num}$ budget therefore requires polynomially many random bits. Additional Bernoulli thinning for the transmission profiles in Sec.~\ref{sec:efficiencies} uses at most $M+N_{\max}$ decisions per sample. An absolute error of $O(\epsilon_{\rm num}/(M+N_{\max}))$ per decision suffices by coupling the choices. These scalar errors can be included in the same numerical budget. The resulting sampler has the claimed polynomial bit complexity.

\subsection{Complete sampling procedure}\label{sec:complete-sampler}
For the supplied interferometer $U$, the sampler in Theorems~\ref{thm:fock} and \ref{thm:gaussian} generates one output configuration in the following three steps. The cutoffs, rejection parameters, and numerical precision are chosen as in the preceding subsections.

\begin{enumerate}
\item \textbf{Choose the photon numbers.}
For Fock state boson sampling, set $N=n$. For Gaussian boson sampling, use $s=M$ and draw $N$ from Eq.~\eqref{eq:countlaw}, conditioned on $N\le N_{\max}$. Then draw $k$ from $\operatorname{Bin}(N,\eta)$. If $k=0$ or $M=1$, output the unique compatible configuration and stop.

\item \textbf{Select the sampling method.}
For Gaussian boson sampling, two branches finish at this step. If $N\le N_0$, draw the lossless output configuration using the prefix sampler of Sec.~\ref{app:prefix}, then select $k$ surviving photons according to Eq.~\eqref{eq:deletion} and output their configuration. If $N>N_0$ and either $N>cM$ or $k>\vartheta N$, output a fixed configuration with total photon number $k$. All remaining sectors, including those of Fock state boson sampling, proceed to step~3.

\item \textbf{Generate a configuration by rejection sampling.}
Propose $x$ uniformly among all $k$-photon configurations. Average $T$ independent estimates from Appendix~\ref{app:evaluation} to obtain $\widehat F_T(x)$, and accept $x$ with probability
\[
 \frac{\clip_{[0,A]}(\widehat F_T(x))}{A}.
\]
Output the first accepted configuration. After each rejection, draw a new candidate and use fresh estimator randomness, for at most $J$ trials in total. If every trial fails, output a fixed configuration with total photon number $k$.
\end{enumerate}

The count bounds give $\epsilon_{\rm count}\le3\epsilon/64$, and finite precision and finite random bits contribute at most $\epsilon_{\rm num}=\epsilon/4$. With the preceding parameter choices, the first four terms in Eq.~\eqref{eq:masterTV} each contribute at most $\epsilon/8$, and the finite trial term contributes at most $\epsilon/(16\mathrm e)$. Thus
\begin{equation}
 \E_U\TV(Q_U,P_U)
 \le\left(\frac{51}{64}+\frac{1}{16\mathrm e}\right)\epsilon
 <0.82\epsilon<\epsilon.
 \label{eq:finalbudget}
\end{equation}
Every branch uses a number of arithmetic operations polynomial in the system size and $1/\epsilon$, uniformly over $0\le r\le r_{\max}$ for the Gaussian input, and Sec.~\ref{app:precision} gives the same bound in bit complexity. This completes the proofs of Theorems~\ref{thm:fock} and \ref{thm:gaussian}.

\section{An arbitrary number of squeezed inputs}\label{sec:vacuum}
We now apply the sampling construction to $s$ equally squeezed inputs and $M-s$ vacuum inputs. The evaluator of Appendix~\ref{app:evaluation} and the count bounds of Appendix~\ref{app:sampling} already apply to this input. The remaining step is to control the correlation tail uniformly in $s$. We first derive the needed reduced purity bound, then combine it with the established sampler.

\subsection{Reduced purity of the squeezed input}\label{app:purity}
Let $\tau_h(s,N)$ be the purity of the normalized $h$-particle reduction of $\ket{\Psi_{s,N}}$, with $N=2n$. Adding vacuum modes preserves this reduced purity. For every $s\ge1$ and $0\le h\le N$, the coefficients in Eq.~\eqref{eq:coreweights} give
\begin{equation}
 D_{s,h}\tau_h(s,N) =\sum_{j=0}^{h}(D_{s,j}+D_{s,j-1})\Gamma_{s,N,j} \times\frac{\binom{h}{j}}{\binom{N}{j}} \frac{(s+N+j-1)!(s+h-1)!} {(s+N-1)!(s+h+j-1)!}.\label{eq:corepurity}
\end{equation}
For $s\ge2$, this identity follows from Eqs.~\eqref{eq:generalsingular} and \eqref{eq:readoutnorm} in the occupied $s$-mode space. For $s=1$, Eq.~\eqref{eq:coreweights} gives $\Gamma_{1,N,0}=1$ and $\Gamma_{1,N,j}=0$ for $j>0$, so both sides equal one. The odd $\Gamma_{s,N,j}$ vanish, and the even weights satisfy Eq.~\eqref{eq:coreone}. These facts imply the uniform bound
\begin{equation}
 \binom Nh\tau_h(s,N)
 \le2\left(1+\frac{2N}{s}\right)^h.\label{eq:taubound}
\end{equation}
To see the resulting estimate explicitly, multiply one summand of Eq.~\eqref{eq:corepurity}, with $\Gamma_{s,N,j}$ omitted, by $\binom Nh/D_{s,h}$. The result is
\begin{equation}
 \binom hj\frac{(N-j)!(s+j-1)!}{(N-h)!(s+h-1)!} \times \frac{(s+N+j-1)!(s+h-1)!}{(s+N-1)!(s+h+j-1)!} \frac{D_{s,j}+D_{s,j-1}}{D_{s,j}}.\label{eq:puritysummand}
\end{equation}
For $c_N=N/s$, the three factors after $\binom hj$ are at most $c_N^{h-j}$, $(1+c_N)^j$, and $2$, respectively. Using $\Gamma_{s,N,j}\le1$ and summing the even $j$ gives
\begin{equation}
 \binom Nh\tau_h
 \le 2\sum_{\substack{j\le h\\j\ {\rm even}}}
     \binom hj c_N^{h-j}(1+c_N)^j
 \le2(1+2c_N)^h.
\end{equation}

\subsection{Conditions for efficient sampling}
\begin{corollary}[Gaussian boson sampling with $s$ squeezed inputs]\label{cor:vacuum}
Fix $r>0$ and a transmission $\eta$ such that
\begin{equation}
 \eta\cosh(2r)<1.\label{eq:region}
\end{equation}
For every $1\le s\le M$, consider the input $\ket{\Psi_{\mathrm G}}$ of Problem~\ref{prob:gaussian}, with $s$ squeezed vacuum states of common squeezing parameter $r$ and $M-s$ vacuum states. For every $0<\epsilon\le1/2$, there is a classical randomized algorithm satisfying
\begin{equation}
 \E_U\TV(Q_U,P^{\mathrm G}_{U,s,r,\eta})\le\epsilon\label{eq:guarantee}
\end{equation}
with worst case running time per sample bounded by $\operatorname{poly}_{r,\eta}(M,1/\epsilon)$ arithmetic operations. The polynomial's degree and constants are independent of $s$.
\end{corollary}

The high probability bound of Eq.~\eqref{eq:typical-accuracy} and the bound for repeated samples in Appendix~\ref{sec:model} also hold in this region. The running time bound can grow as $\eta\cosh(2r)$ approaches one.

\begin{proof}
For $s<M$, the state $\ket{\Psi_{s,N}}$ occupies an $s$ dimensional single particle subspace before the interferometer. Its Haar weights $W_h(k)$ nevertheless depend on the ambient mode count $M$. For example, the first order has weight
\begin{equation}
 W_1(k)=\frac{k(M-s)}{s(M+k)},\label{eq:firstorder}
\end{equation}
which is positive for $k>0$ and $s<M$. Vacuum embedding thus introduces odd correlation orders that vanish when every input mode is squeezed.

The reduced purity bound in Eq.~\eqref{eq:taubound}, combined with Eq.~\eqref{eq:puritybound}, yields
\begin{equation}
 W_h(k)\le2\left[\frac{k}{N}
                  \left(1+\frac{2N}{s}\right)\right]^h.\label{eq:vacweight}
\end{equation}
This step removes any dependence on the vacuum fraction. Choose constants
\begin{equation}
 c>\sinh^2r,\quad\vartheta>\eta,
 \quad\tilde\chi=\vartheta(1+2c)<1.\label{eq:vacgood}
\end{equation}
Such constants exist precisely when Eq.~\eqref{eq:region} holds, because $\cosh(2r)=1+2\sinh^2r$. On $N\le cs$ and $k\le\vartheta N$,
\begin{align}
 \E_U\|F_U-F_U^{\le L}\|_2^2
 &\le\frac{2\tilde\chi^{L+1}}{1-\tilde\chi},\label{eq:vactail}\\
 \E_U\|F_U\|_2^2&\le\frac{1+\tilde\chi}{1-\tilde\chi}.\label{eq:vacsecond}
\end{align}
These bounds include the odd correlation orders introduced by vacuum inputs and all output configurations with collisions. For small $s$, the event $N>cs$ can carry appreciable probability. We therefore use prefix sampling in every sector with $N\le N_0=O_{r,\eta}(1+\log(1/\epsilon))$ and apply the geometric bound only in the stated region with $N>N_0$. Equations~\eqref{eq:densetail} and \eqref{eq:binomialtail} control the remaining probability uniformly in $s$, as shown in Sec.~\ref{app:counts}.

The Gaussian evaluator of Appendix~\ref{app:evaluation} applies to the same $s$-mode input sector. Together with the count bounds of Sec.~\ref{app:counts}, Eqs.~\eqref{eq:vactail} and \eqref{eq:vacsecond} give $C_2=O_{r,\eta}(1)$ and a geometric tail $S_L$. Substituting them into Eq.~\eqref{eq:masterTV} and using the finite precision analysis of Sec.~\ref{app:precision} proves the stated error and running time uniformly in $s$.
\end{proof}

The optical condition can also be written
\begin{equation}
 \eta\left(1+2\frac{\overline N_{\mathrm{in}}}{s}\right)<1,
 \qquad \overline N_{\mathrm{in}}=s\sinh^2r.\label{eq:occupancyform}
\end{equation}
The sufficient condition also bounds the mean detected photon number per squeezed input as $\eta\sinh^2r<(1-\eta)/2$. For example, $r=\frac12\log3$ gives $\sinh^2r=1/3$ and requires $\eta<3/5$. This corresponds to about $4.77$ dB of input squeezing and a mean input photon number $s/3$.

\section{Scope of the sampling results}\label{app:extensions}

\subsection{Dependence on Haar averaging}\label{sec:limits}
The Haar average enters our analysis in two places. First, the correlation weights $W_h(k)$ of Appendix~\ref{app:correlations} are Haar averages, so the tail beyond the cutoff is controlled only on average over interferometers. Second, the variance of the random estimate in Lemma~\ref{lem:mask} is also bounded only on average over interferometers. The running time bound holds for every interferometer, but the accuracy bound does not. Through Eq.~\eqref{eq:typical-accuracy}, the average error also yields a bound that holds with high probability for a circuit drawn once.

For a specified interferometer, such as the structured interferometers discussed in Sec.~\ref{sec:main-results}, separate bounds on both the correlation tail and the estimator variance are therefore needed.

\subsection{Mode-dependent transmissions}\label{sec:efficiencies}
Theorems~\ref{thm:fock} and \ref{thm:gaussian} also cover unequal transmissions in the final output channels when they satisfy $0\le\eta_i\le\eta_*$ for a fixed $0<\eta_*<1$. Generate a sample $x$ using uniform transmission $\eta_*$, then independently draw the count in mode $i$ from $\operatorname{Bin}(x_i,\eta_i/\eta_*)$. Pure-loss channels compose by multiplying their transmissions. This classical thinning therefore gives the desired output law, and total variation error cannot increase. The transmission profile is supplied at the sufficient finite binary precision for the requested error. A measured average transmission alone does not specify the upper bound $\eta_*$.

Independent source loss can also be included for single-photon inputs. If a fixed uniform loss is already present, first select which occupied input modes survive the additional source loss, then apply Theorem~\ref{thm:fock} to the selected input subset. When source loss is the only loss, suppose its transmissions satisfy $p_i\le\eta_*<1$ for a fixed $\eta_*$. Select input $i$ with probability $p_i/\eta_*$ and use uniform transmission $\eta_*$ in the sampler. Conditional on the selected subset, a permutation of the input columns gives the standard single-photon configuration. For a source profile independent of $U$, Haar invariance under these permutations and convexity of total variation preserve the mean error bound after averaging over the subset. Selecting a subset has polynomial cost and requires one sampler call, including the vacuum case.

General path-dependent loss inside the interferometer requires additional analysis.

\subsection{Comparison with other simulation methods}
Section~\ref{sec:previous} summarizes these methods. Here we record the scalings behind that summary and compare the assumptions on the interferometer.

For single-photon inputs, the classical approximation of the lossy state applies when the number of surviving photons grows more slowly than the square root of the number of input photons~\cite{OszmaniecBrod2018}, and thermal noise approximations cover architectures whose transmission decays exponentially with circuit depth~\cite{GarciaPatron2019}. For Gaussian inputs at fixed squeezing, the error bounds $O_r(\sqrt{s}\,\eta)$ of the Gaussian approximation~\cite{Qi} and $O_r(s\eta^3)$ of the reset method~\cite{LimReset} give the sufficient scalings $\eta=O(s^{-1/2})$ and $\eta=O(s^{-1/3})$, which permit mean detected photon numbers of order $s^{1/2}$ and $s^{2/3}$. In all these cases the fraction of input photons that survives tends to zero. The loss analyses of tensor network methods identify the same square root scale, with logarithmic refinements in the entanglement bounds~\cite{OhNP,OhMPS2026}. Increasing the bond dimension can reduce the error, but at $\eta=\Theta(s^{-1/2})$ the Gaussian analysis does not establish a cost polynomial in both system size and $1/\epsilon$~\cite{OhNP}.

Interference truncation~\cite{Renema} evaluates a probability truncated at order $L$ through $M^{O(L)}$ auxiliary configurations and draws samples with a Metropolis chain. Our construction uses the same physical suppression of high-order correlations but replaces both steps. The two-sided projection and random grouping of Appendix~\ref{app:evaluation} avoid the $M^{O(L)}$ enumeration, with cost polynomial in system size and singly exponential in $L$ and $\log(1/\epsilon)$, and the rejection sampler of Sec.~\ref{sec:sampler} stops after a bounded number of trials with an explicit total variation bound.

The interferometer assumptions also differ. Our accuracy bound is averaged over Haar circuits, while the approximations for single-photon inputs and the Gaussian approximation and reset bounds hold for every $U$, and tensor network methods apply to specified circuits with cost controlled by their entanglement.

\section{Comparison with sampling hardness}\label{app:hardness}

\subsection{Fock state boson sampling}\label{sec:hardness}
Conditioning on the event that no photon is lost~\cite{AaronsonBrod} yields conditional hardness when the mean loss is logarithmic~\cite{GoDistinguishability}. For $n$ single photons in distinct input modes, write $P_U^{(\eta)}=P^{\mathrm F}_{U,n,\eta}$ for the full lossy output distribution. For every $U$,
\begin{equation}
 P_U^{(\eta)}(x)=\eta^nP_U^{(1)}(x),\qquad |x|=n.
 \label{eq:nolossbranch}
\end{equation}
The event $E=\{x:|x|=n\}$ therefore has probability $p=\eta^n$, independent of $U$, and conditioning on $E$ gives the ideal distribution. If the mean number of lost photons obeys
\begin{equation}
 (1-\eta)n=O(\log n),
 \label{eq:logloss}
\end{equation}
then $-\log p=-n\log\eta=O(\log n)$, so $p\ge n^{-c}$ for some fixed $c>0$.

For an approximate lossy sampler with output law $Q_U$, let $\delta_U=\TV(P_U^{(\eta)},Q_U)$. Whenever $Q_U(E)>0$, conditioning gives
\[
 \TV\bigl(P_U^{(1)},Q_U(\cdot\mid E)\bigr)\le\delta_U/p.
\]
Draw from $Q_U$ until an outcome in $E$ is obtained, with a limit of $R$ independent trials. If every trial fails, output a fixed $n$-photon configuration. For $\delta_U\le p/2$, the acceptance probability is at least $p/2$ and the failure probability is at most $\mathrm e^{-pR/2}$. By Markov's inequality, the remaining interferometers have Haar probability at most $2\E_U\delta_U/p$. The output law $\widehat Q_U$ of this bounded procedure consequently satisfies
\begin{equation}
 \E_U\TV\bigl(P_U^{(1)},\widehat Q_U\bigr)
 \le\frac{3\E_U\delta_U}{p}+\mathrm e^{-pR/2}.
 \label{eq:noloss-error}
\end{equation}
Taking $\E_U\delta_U\le p\epsilon/6$ and $R=\lceil(2/p)\log(2/\epsilon)\rceil$ gives mean error at most $\epsilon$. Under Eq.~\eqref{eq:logloss}, an inverse polynomial target error requires only inverse polynomial lossy error and polynomially many trials. When $M=\operatorname{poly}(n)$, the hardness conclusion therefore inherits the ideal sampling assumptions in Refs.~\cite{AaronsonArkhipov,Bouland}, in their respective mode scalings.

At fixed $\eta$, the probability $p$ decays exponentially with $n$, so recovering ideal samples this way requires exponentially many trials. In this regime, the sampler of Theorem~\ref{thm:fock} still samples the full lossy distribution efficiently. Its runtime bound depends on $\eta$, as seen from the required correlation cutoff.

The sampler of Theorem~\ref{thm:fock} uses the cutoff in Eq.~\eqref{eq:lowlosscut}, which keeps the mean squared error from orders above $L$ below $(\epsilon/16)^2$. For $\epsilon=n^{-a}$ with fixed $a>0$, the right side of Eq.~\eqref{eq:lowlosscut} is $O(\log n)$ at fixed $\eta$. It becomes $\Theta(n)$ when $1-\eta=\Theta(\log n/n)$. The evaluation cost is exponential in $L$, so our bound on the running time is no longer polynomial in the regime of logarithmic loss. The intermediate range $\log n\ll(1-\eta)n\ll n$ remains open.

\subsection{Gaussian boson sampling}
Conditional hardness under loss has been studied for equal squeezing in all $M$ input modes with a Haar random interferometer~\cite{GoLoss}. The analysis uses the dilute scaling $M=\Theta(n^\gamma)$ for sufficiently large $\gamma>2$, where $n$ now denotes the mean detected photon number. The squeezing is chosen through
\begin{equation}
 n=\eta M\sinh^2r.
 \label{eq:gbs-hardness-squeezing}
\end{equation}
Its hardness result concerns logarithmic mean loss, and its reduction for estimating probabilities uses a range of transmissions. The sampling hardness statement at a specified transmission additionally assumes that increasing loss monotonically reduces simulation complexity at fixed total variation error, alongside the ideal average case hardness conjecture. In this regime of low loss, the specified photon count occurs with inverse polynomial probability, so the hardness conclusion also applies to the full output distribution.

At any fixed $\eta$, the same dilute scaling has $r\to0$ and is covered by our sampling construction. Since the bound of Theorem~\ref{thm:gaussian} applies to every smaller squeezing parameter, taking any fixed upper value of the squeezing covers this regime, with the bounds on photon counts and source precision proved in Secs.~\ref{app:counts} and \ref{app:precision}. The resulting cost remains polynomial in $M$ and $1/\epsilon$, hence in $n$ and $1/\epsilon$ when $M=\operatorname{poly}(n)$. The hardness regime of logarithmic loss has $\eta\to1$, where our runtime bound at fixed transmission does not apply.

For $s$ equally squeezed inputs and $M-s$ vacuum inputs, Corollary~\ref{cor:vacuum} establishes efficient sampling under the sufficient condition in Eq.~\eqref{eq:occupancyform}. At fixed $\eta$, this condition is eventually satisfied whenever $\overline N_{\mathrm{in}}/s\to0$, even if $\overline N_{\mathrm{in}}\to\infty$.

The hardness result for dense bipartite Gaussian boson sampling concerns pairs of modes prepared in two-mode squeezed states and mixed by separate interferometers on the two arms~\cite{Bouland}. The argument uses a Haar unitary relating the arms and assumptions on the resulting permanent estimation problem. The error bound of Theorem~\ref{thm:gaussian} is averaged over Haar unitaries that mix all modes, so it does not establish efficient sampling for the bipartite interferometer ensemble.

A separate result concerns the computation of probabilities rather than sampling. For local noise below a threshold, including photon loss, computing output probabilities of Gaussian boson sampling with Haar random interferometers to exponentially small additive error remains \#P-hard on average~\cite{Deshpande2022}. This hardness of individual probabilities does not imply hardness of approximate sampling and is therefore compatible with the efficient samplers established here.

\bibliography{references}
\end{document}